\documentclass[twocolumn,nofootinbib]{revtex4-2}

\pdfoutput=1
\usepackage{graphicx}
\usepackage{bm}
\usepackage{xcolor}
\usepackage{amsmath,amssymb,amsfonts}
\usepackage{epstopdf,epsfig}
\usepackage{multirow}
\usepackage[bbgreekl]{mathbbol}
\usepackage{hyperref}

\graphicspath{{figures/}}

\makeatletter
\def\amsbb{\use@mathgroup \M@U \symAMSb}
\makeatother

\newcommand{\be}{\begin{equation}}
\newcommand{\ee}{\end{equation}}
\newcommand{\bea}{\begin{eqnarray}}
\newcommand{\eea}{\end{eqnarray}}
\renewcommand{\vec}[1]{{\bf #1}}
\newcommand{\mnras}{MNRAS}
\newcommand{\aap}{Astronomy \& Astrophysics}
\begin{document}

\title{Clusters of heavy particles in two-dimensional Keplerian turbulence}
\author{Fabiola A. Gerosa}\email{fabiola.gerosa@oca.eu}
\author{H\'elo\"ise M\'eheut}
\affiliation{Universit\'e C\^ote d'Azur, Observatoire de la C\^ote d'Azur, CNRS, Laboratoire Lagrange, Nice, France}
\author{J\'er\'emie Bec}
\affiliation{Universit\'e C\^ote d'Azur, Inria, CNRS, Sophia-Antipolis, France\\Mines Paris, PSL University, CNRS, Cemef, Sophia-Antipolis, France}

\begin{abstract}
Protoplanetary disks are gaseous systems in Keplerian rotation around young stars, known to be turbulent. They include a small fraction of dust from which planets form. In the incremental scenario for planet growth, the formation of kilometer-size objects (planetesimals) from pebbles is a major open question. Clustering of particles is necessary for solids to undergo a local gravitational collapse.\\
To address this question, the dynamic of inertial particles in turbulent flows with Keplerian rotation and shear is studied. Two-dimensional direct numerical simulations are performed to explore systematically two physical parameters: the rotation rate, which depends on the distance to the star, and the particle response time, which relates to their size. \\
Shear is found to drastically affect the characteristics of the turbulent flow destroying cyclones and favoring the survival of anticyclones. 
Faster Keplerian rotation enhances clustering of particles. For intermediate sizes, particles concentrate in anticyclones. These clusters form in a hierarchical manner and merge together with time. For other parameter values, solids concentrate on fractal sets that get more singular with rotation. The mass distribution of particles is then found to be multifractal with small dimensions at large orders, intriguing for triggering their gravitational collapse. Such results are promising for a precise description and better understanding of planetesimal formation.
\end{abstract}

\maketitle

\section{Introduction}

A stellar system appears after the gravitational collapse of a molecular cloud. While most matter accumulates in the young star, a tiny fraction (less than a few percents) sediments in a protoplanetary disk, which rotates around the star and serves as a stage for the formation of planets, moons, asteroids, and other small bodies. Such disks are mainly composed of gas, but also contain some sub-micrometer dust particles ($\approx 1\%$ in mass) that are the building blocks of larger solid bodies. One of the major open questions in the theory of planet formation is the emergence of planetesimals (objects with sizes of the order of kilometers) that are subsequently able to further accrete solids and even capture the surrounding gas. Indeed classical scenarios, in which dust particles grow in a hierarchical, sequential manner~\cite{safronov1972evolution}, are facing several difficulties.

A first issue relates to the outcomes of collisions between solids~\cite{blum1999aspects}. At sizes larger than centimeters,\footnote{The sizes given in this paragraph vary with distance to the star. Examples given here correspond to 1 Astronomical Unit.} collisions between pebbles are expected to involve too violent impact velocities. They can hardly lead to coagulation, but rather result in bouncing or fragmentation, making hierarchical growth by far too slow and inefficient at such sizes. A second issue comes from the radial drift that is experienced by solid particles~\cite{weidenschilling77}. Protoplanetary disks are indeed expected to have a large-scale radial pressure gradient, so that the gas rotates with a slightly sub-Keplerian speed.  Solid particles are insensitive to this pressure gradient and rotate with the exact Keplerian velocity. They are thus dragged by the gas that slows them down, making them drift radially toward the central star. This process occurs on too short time scales to form planetesimals by sequential growth. 

Drift and fragmentation make the growth of planetesimals difficult. Nevertheless, it is generally assumed that such effects can be overcome by a direct self-gravitational collapse of dust particles that can bypass hierarchical coalescence. This requires both gravitational interactions to be dominant over the dispersion induced by the Keplerian shear, and velocity differences between dust grains to be sufficiently feeble. This could occur during the settling of solids in the disk's equatorial plane~\cite{goldreich1973formation}. We focus here on another scenario involving spatial inhomogeneities, whose presence provides favorable conditions. Self-gravity can indeed become dominant in regions where the density of solids is large and their velocity dispersion narrow~\citep{GSL20}. In such a scenario a major role is played by the turbulence of the carrier gas~\citep{IKE18,HC20}. It is even largely admitted that the back reaction of the particles on the gas could itself trigger turbulence. Underlying mechanisms involve the streaming instability~\citep{johansen2007protoplanetary} or other resonant drag instabilities, in which the decoupling between gas and particles (through settling, drift, buoyancy, or magnetic waves) strongly amplifies tiny fluctuations in the concentration of solids~\citep{SH18-2}. Recent evidences have however been obtained that, on the one hand, an already developed turbulent state impedes the growth of such instabilities~\cite{chen2020efficient,umurhan2020streaming} and that, on the other hand, they prohibitively require starting from already large particle sizes~\cite{carrera2022streaming}.

Besides planet formation, the origin and nature of turbulence in protoplanetary disks, and how it couples to particle dynamics, is key to several other astrophysical questions. For instance, disks are known to have a rather short lifetime (of the order of a few millions years). The dissipation of angular momentum by the gas molecular viscosity is orders of magnitude too low to explain this fast accretion onto the central star. Yet turbulence is an excellent candidate to transport matter on shorter timescales, hence providing estimates on the expected turbulent intensity of disks.  Transport mechanisms in protoplanetary disks, possible instabilities and the development of a turbulent state are still the topic of intense work (see, \textit{e.g.}, \cite{armitage2011dynamics,LEF22} for a review). Whether it originates from magneto-rotational effects or purely compressible or incompressible hydrodynamical instabilities, disk turbulence is confined, in rotation, stratified, and thus generally displays fairly strong two-dimensional features, as for instance long-lived vortical structures~\cite{bracco1998spotted}.

The presence of such vortices might clearly impact dust accretion and consequently the formation of planetesimals.  While heavy inertial particles are usually expelled from rotating regions by Maxey's centrifugal effect~\cite{maxey1987gravitational}, the presence of shear and rotation in protoplanetary disks actually has an opposite influence. Numerical~\cite{TANGA1996158,bracco1999particle}, analytical~\cite{barge1995,CHA00} and experimental~\cite{Solomon1993} studies indeed give evidence that dust particles cluster in some of the structures of the flow. More specifically, solid particles experience a Coriolis force that can either accelerate their ejection from cyclones (vortices with the same sign as global rotation) or, when strong enough, overcome centrifugal forces in anticyclones. This pushes the particles toward the core of anticyclonic vortices, possibly creating extremely dense point clusters that are excellent candidates for gravitational instabilities. These mechanisms have been known for years, but their study remained essentially qualitative or limited to model flows. Understanding further what is the impact of vortex clustering on planetesimal formation requires more quantitative insights, such as estimates on involved timescales in the presence of turbulent fluctuations, dependence on physical parameters, nature of the associated mass distribution, etc. The present work aims providing insights on such questions.

We present in this paper 2D direct numerical simulations with the idea of shedding some lights on the complex problem of dust dynamics in turbulent protoplanetary disks. We investigate the clustering properties of solid particles in a forced, developed turbulent incompressible flow subject to Keplerian rotation and shear. We follow the ``shearing box'' approach, which consists in solving locally the Navier--Stokes equation with a mean constant shear and with periodic boundary conditions on a domain that is distorted in the direction of the mean velocity. A number of simulations are performed varying systematically the two main physical parameters of the problem: the particle response time $\tau_{\rm p}$, which measures their inertia and their lag on the gas flow, and the rotation rate $\Omega$, which prescribes the mean shear.  We then analyse quantitatively various dynamical and statistical features of solid particles, and in particular their clustering properties when flow structures, typical of a turbulent flow in rotation, are present. Tools borrowed from the study of dynamical systems, such as Lyapunov exponents and fractal dimensions, are used to characterize and quantify particle clusters. 

Before focusing on particles, we find that the mean shear has noticeable impacts on the properties of the turbulent flow. Both energy and enstrophy budgets, as well as energy and enstrophy spectra, change substantially with increasing rotation. Strong rotation also leads to pronounced anisotropies in the flow and thus to increased skewness and kurtosis of the vorticity distribution. Shear is found to cause the preferential formation and survival of anticyclonic vortices. Concerning particles, we find that at specific values of the physical parameters, they can form point clusters located inside anticyclones. This strong clustering requires large-enough values of the rotation rate $\Omega$ and intermediate values of the particle response time $\tau_{\rm p}$. These clusters are found to form in a hierarchical manner and to merge one with the other when time increases. We propose a simple kinetic model that catches most aspects of their evolution and distribution. Outside this extreme regime, solid particles can nevertheless concentrate on dynamically evolving fractal sets whose dimensions non-trivially depend on the rotation rate. Evidence is obtained that the mass distribution of particles is then multifractal, with a dimension that decreases and saturates at large orders. Such a behavior is key to quantify the probability to get a large local density of solids and to trigger self-gravitating instabilities.  In the strong clustering regime, such instabilities would occur during transients. 

This paper is organised as follows. We first introduce and characterise in section \ref{section:shearingbox} the flow used for the numerical simulations and we explain in section \ref{section:particles} the considered dynamics of particles. In Section \ref{section:fractvsstrong} we show the results of the simulations, highlighting the different behaviors observed for the distribution of inertial particles. In section \ref{section:fract} and section \ref{section:strong} the mass distribution of particles is inspected in the regime of fractal and strong clustering, respectively. Finally, in section \ref{section:disc} we conclude the paper summarizing the results of this study and connecting it back to the astrophysical context.

\section{The two-dimensional shearing box}
\label{section:shearingbox}

\begin{figure}[htpb]
    \centering
    \includegraphics[width=\columnwidth]{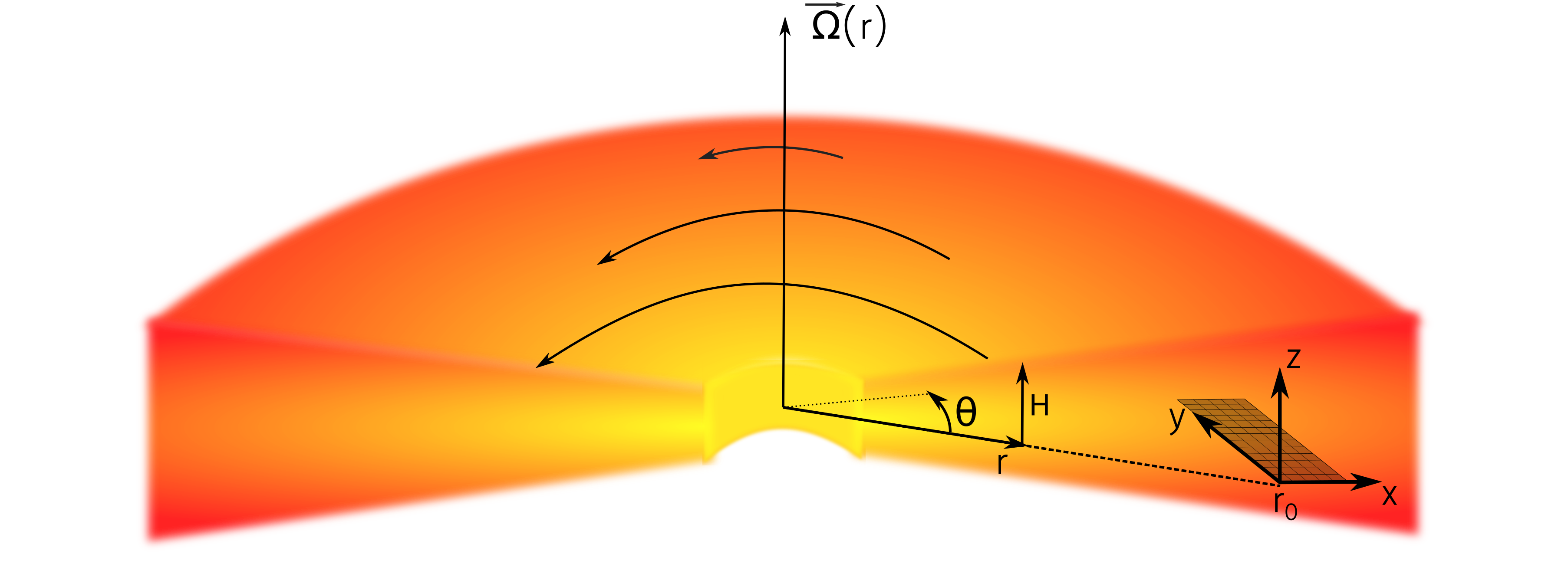}
    \caption{Schematic view of a protoplanetary disk and of the rotating shearing box that we here consider.}
    \label{fig:3Ddisk}
\end{figure}

Dust and gas dynamics in protoplanetary disks, as in most astrophysical situations, involve gravity and rotation. The balance between gravity and centrifugal forces defines the Keplerian angular velocity, which is given by
\be
\Omega(r) = \sqrt{{\mathcal{G}\,{M}_\star}/{r^3}}
\ee
where $M_\star$ is the mass of the central star, $\mathcal{G}$ is the gravitational constant, and $r$ is the distance to the central star. This velocity profile corresponds to a rotating flow with a strong shear (see Fig.~\ref{fig:3Ddisk}).

\subsection{Model and equations of motion}
\label{subection:model}
We focus on a small box whose edge is located at a distance $r_0$ from the star and which is rotating with the reference angular velocity $\Omega(r_0)$ (see Fig.~\ref{fig:3Ddisk}). We then write the dynamics in the co-moving coordinates $x = r-r_0$ and $y = r_0\,(\theta - \Omega(r_0)t)$, where $\theta$ denotes the azimuthal angle. When the reference distance $r_0$ is much larger that the box size, curvature terms can be neglected. Local variations of the Keplerian angular velocity are shearing our small box. We indeed have $\Omega(r) - \Omega(r_0) \approx (r-r_0)\,\tfrac{\mathrm{d}}{\mathrm{d}r}\Omega(r_0) = -\tfrac{3}{2}\,\Omega(r_0)\,x/r_0$.
In the co-rotating frame, gravity and centrifugal terms balance at $r=r_0$. Away from $r_0$, a tidal force and a Coriolis term must be added. This model, introduced by Hill~\citep{H78} in the late 70's, leads to write the momentum conservation for the gas as:
\be
 \partial_t \vec{v} + \vec{v}\cdot\vec{\nabla} \vec{v} = - \frac{1}{\rho_{\rm g}}\vec{\nabla} p + \nu \vec{\nabla}^2 \vec{v} -2\vec{\Omega}\times \vec{v} + 3\Omega^2\, x\, \vec{e}_x,
 \label{eq:NS}
\ee
where $\vec{v}$ is the total gas velocity in the co-moving frame. $\rho_{\rm g}$ and $\nu$ are its mass density and kinematic viscosity, respectively. We have here denoted $\Omega = \Omega(r_0) = {\rm const}$ and $\vec{\Omega} = \Omega\,\vec{e}_z$. For an incompressible flow, the equation of continuity reads:
\be
  \vec{\nabla} \cdot \vec{v}=0,
\ee
which hence prescribes the pressure $p$ in \eqref{eq:NS}. The use of incompressibility is justified by the fact that, at the small scales that we consider, density waves are less important. Therefore, density can be assumed constant over time. Note that averaging~\eqref{eq:NS} leads to the mean shear $\langle \partial_x \textrm{v}_y \rangle = -\tfrac{3}{2}\,\Omega$. 

We focus here on two-dimensional flows. Several considerations partly legitimize this choice. A first idea comes from the limit of small Rossby numbers $Ro = U/(L\Omega)\ll 1$, where $U$ and $L$ are characteristic velocity and length scales of the flow. The Taylor–Proudman theorem \citep{P16,T17} ensures that any motion that occurs on timescales longer than $\Omega^{-1}$ becomes independent of the $z$ coordinate. The flow is then approximately two-dimensional and characterized by the well-known Taylor columns \citep{T22}, that are long columnar eddies oriented parallel to the rotation axis.  A second idea comes from the fact that protoplanetary disks are expected to be highly stratified in the vertical direction $z$. This could be another mechanism by which the gas dynamics becomes almost two-dimensional. In this case, rather than being long and columnar, eddies are now flat with the shape of pancakes. Their formation originates then from internal gravity waves with group velocity perpendicular to the gravitational force~\cite{davidson2015turbulence}. Both kinds of structures, columns and pancakes, typically coexist in the presence of both rotation and stratification~\citep{C01}. However, the global dynamics of protoplanetary disks actually involve baroclinic amplifications, thermal transfers, radial stratification, elliptical instabilities, etc., that are all affecting the formation and survival of coherent eddies~\cite{barge2016vortices}. Studying such instabilities is beyond the scope of our work. We rather focus on two-dimensional solutions of \eqref{eq:NS} with turbulent fluctuations and vortex dynamics maintained in a developed, statistically steady state thanks to the addition of an external random forcing.

In our settings, the fluid velocity solves the two-dimensional incompressible Navier--Stokes equation. The divergence-free turbulent fluctuations $\vec{u} = \vec{v}+\tfrac{3}{2}\,\Omega\,x\,\vec{e}_y$ are such that the associated vorticity $\omega = \nabla^{\mathsf{T}}\!\cdot\vec{u} \equiv \partial_x \textrm{u}_y- \partial_y \textrm{u}_x$ solves
\begin{equation}
  \partial_t \omega + \vec{v}\cdot\nabla \omega =  \nu\nabla^2 \omega
  -\alpha\,\omega+f_{\omega}.
  \label{eq:vorticity}
\end{equation}
Note that, because of incompressibility, the Coriolis force does not contribute to the vorticity dynamics. We are here applying a stochastic forcing $f_\omega$ in order to maintain a developed turbulent state. This forcing is assumed Gaussian, with zero mean, homogeneous, isotropic, white in time, and with spatial correlations concentrated at large scales. As common in two-dimensional turbulence, we suppose that vorticity experiences a linear friction with coefficient $\alpha$. Such a term prevents kinetic energy from piling up in the flow. It typically originates, either from the friction between different layers in the stratified case, or from the effect of boundaries when the two-dimensional effective dynamics corresponds to a vertical average.

\subsection{Direct enstrophy cascade}
\begin{figure}[htpb]
    \centering
    \hspace{-0.5cm}
    \includegraphics[width=.5\textwidth]{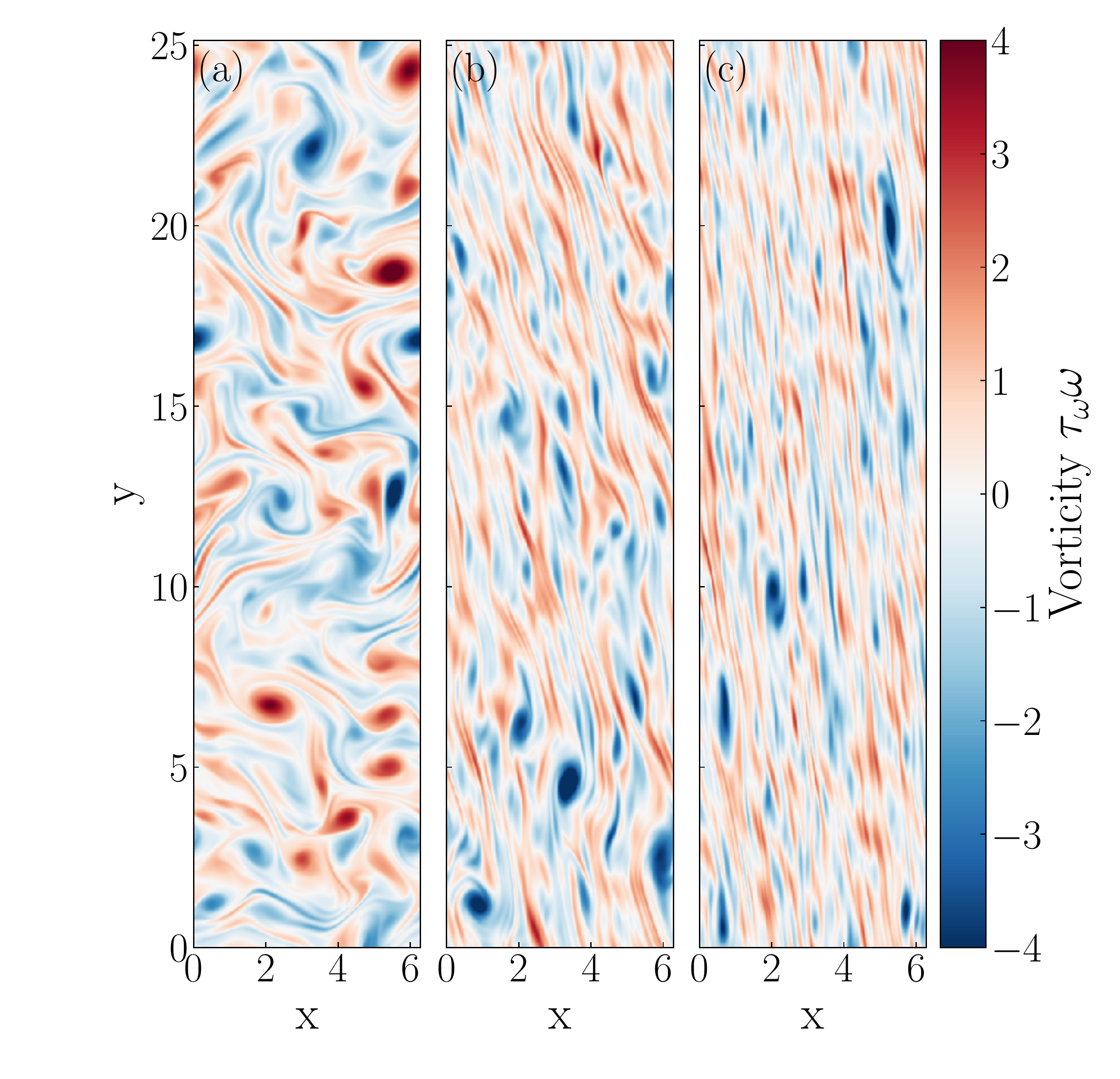}
    \caption{Three different snapshots of the gas vorticity field obtained in the absence of rotation and shear (a), for an intermediate value of $\Omega$ (b), and for a large value of $\Omega$ (c).}
    \label{fig:flow}
\end{figure}

Figures~\ref{fig:flow}a, b, and c show snapshots of the gas vorticity field (normalized by $\tau_{\omega}^{-1}$) for various values of the rotation frequency $\Omega$. One clearly observes, at a qualitative level, that shear sustains anticyclonic vortices (rotating in the opposite direction of $\vec{\Omega}$, \textit{i.e.} with a negative sign, in blue). Because the space average of the fluctuating vorticity $\omega$ is conserved and equal to $0$, positive values (in red) distribute in the background as denser and denser filaments. Another observation is that shear tends to stretch and align anticyclones with the direction of the mean flow. At large values of the rotation, structures typically look like ellipses. We will turn back later to estimate how their aspect ratio depends on the shear rate.

The developed turbulent state that we observed in Fig.~\ref{fig:flow} is characterized by two inviscid quadratic invariants, namely the kinetic energy $\mathcal{E} = \tfrac{1}{2}\langle\|\vec{u}\|^2\rangle$ and the enstrophy $\mathcal{Z} = \tfrac{1}{2}\langle\omega^2\rangle$, whose budgets satisfy
\begin{eqnarray}
  \frac{\mathrm{d}\mathcal{E}}{\mathrm{d}t} = 0 &=& -2\nu\,\mathcal{Z} - 2\alpha\,\mathcal{E} -\tfrac{3}{2}\,\Omega\,\mathcal{R} + \varepsilon_{\rm I}, \label{eq:energy_budget}\\
  \frac{\mathrm{d}\mathcal{Z}}{\mathrm{d}t} = 0 &=& -2\nu\,\mathcal{P} - 2\alpha\,\mathcal{Z} + \eta_{\rm I}. \label{eq:enstrophy_budget}
\end{eqnarray}
We have here introduced the Reynolds shear stress $\mathcal{R} = \langle u_x\,u_y\rangle$ and the palinstrophy $\mathcal{P} = (1/2)\langle\|\nabla\omega\|^2\rangle$. The quantities $\varepsilon_{\rm I}$ and $\eta_{\rm I}$ are the injection rates of kinetic energy and enstrophy, respectively, and are fully determined by the stochastic forcing. Because of the absence of vortex stretching in 2D, the only source of enstrophy in (\ref{eq:enstrophy_budget}) is forcing. This implies that, contrarily to 3D, energy is not dissipated by viscosity but is transferred to large scales by an inverse cascade (see, e.g., \cite{boffetta2012two}). Still, vorticity gradients, and thus palinstrophy, can become very large in 2D, leading to a direct cascade of enstrophy.

The two injection rates define a forcing length scale $\ell_f = (\varepsilon_{\rm I}/\eta_{\rm I})^{1/2}$. Similarly, by balancing the different dissipative rates in (\ref{eq:energy_budget}) and (\ref{eq:enstrophy_budget}), one can introduce a friction scale $\ell_\alpha = (\mathcal{E}/\mathcal{Z})^{1/2}$ and a viscous dissipation scale $\ell_\nu = (\mathcal{Z}/\mathcal{P})^{1/2}$. These length scales define two Reynolds numbers: an outer-scale Reynolds number $R_\alpha = \ell_\alpha/\ell_f$ which measures the ratio between inertial and frictional forces, and a viscous Reynolds number $R_\nu = (\ell_f/\ell_\nu)^2$, which balances inertial and viscous forces. These two numbers prescribe the extensions of the inverse energy cascade and of the direct enstrophy cascade, respectively. We here focus on the direct cascade of enstrophy and we prescribe $\ell_\alpha \gtrsim \ell_f \gg \ell_\nu$, so that $R_\alpha$ is of the order of 1 and $R_\nu\gg 1$.

As seen from Fig.~\ref{fig:flow}, a mean shear ($\Omega \neq 0$) develops anisotropies in the flow and the energy budget is affected by the Reynolds stress. The importance of shear has to be measured by non-dimensionalizing it with a characteristic time scale of the flow. Still, the flow timescales are themselves modified by shear, so this choice cannot be made a priori. In our protocol, the only time scale that is prescribed by the simulation setup is the forcing time scale $\tau_{\rm f} = (\ell_f^2/\varepsilon_I)^{1/3}= \eta_{\rm I}^{-1/3}$. The influence of shear on the energy and enstrophy budget is then measured by the non-dimensional shear rate parameter $\sigma = \tfrac{3}{2}\,\Omega\,\tau_{\rm f}$. The three illustrating values shown in Fig.~\ref{fig:flow} are (a) $\sigma = 0$, (b) $\sigma= 5.7$, and (c) $\sigma=14.2$. In the developed regime attained once $\sigma$ has been prescribed, the shear needs to be compared to the typical dynamical timescale of the direct cascade, namely $\tau_{\omega} = (2\mathcal{Z})^{-1/2}=\langle\omega^2\rangle^{-1/2}$. 

Direct numerical simulations are performed using a pseudo-spectral solver. To construct periodic solutions that account for the mean flow, we follow \cite{rogallo1981numerical,rogers1987structure,HGB95,pumir1996turbulence} and integrate (\ref{eq:vorticity}) on a distorting frame defined by $x' = x$, $y'=y+\tfrac{3}{2}\Omega\,t\,x$. The integration domain is the periodic rectangle $(x',y')\in[0,2\pi[\times[0,8\pi[$, with a sufficiently large aspect ratio in order to limit spurious geometrical effects at large shears. The distorted grid is regularly shifted back to the Cartesian grid at times multiple of the shear time $\tfrac{2}{3}\Omega^{-1}$. Such methods are now standard for local numerical simulations of astrophysical disks~\citep{UR04,LL05} where they are coined as the ``shearing box'' approach.
In our two-dimensional settings, we make use of the vorticity formulation (\ref{eq:vorticity}), together with the Biot--Savard law to obtain the fluctuating velocity $\vec{u}$ as a function of the vorticity $\omega$. The stochastic forcing is approximated as shot noise with spatial power spectrum concentrated over wavenumbers $\|\bm k\|=4$. Time marching uses a second-order Runge--Kutta method, which is explicit for the non-linear term and implicit for the friction and viscous terms.

\begin{figure}[htpb]
    \centering
    \includegraphics[width=.45\textwidth]{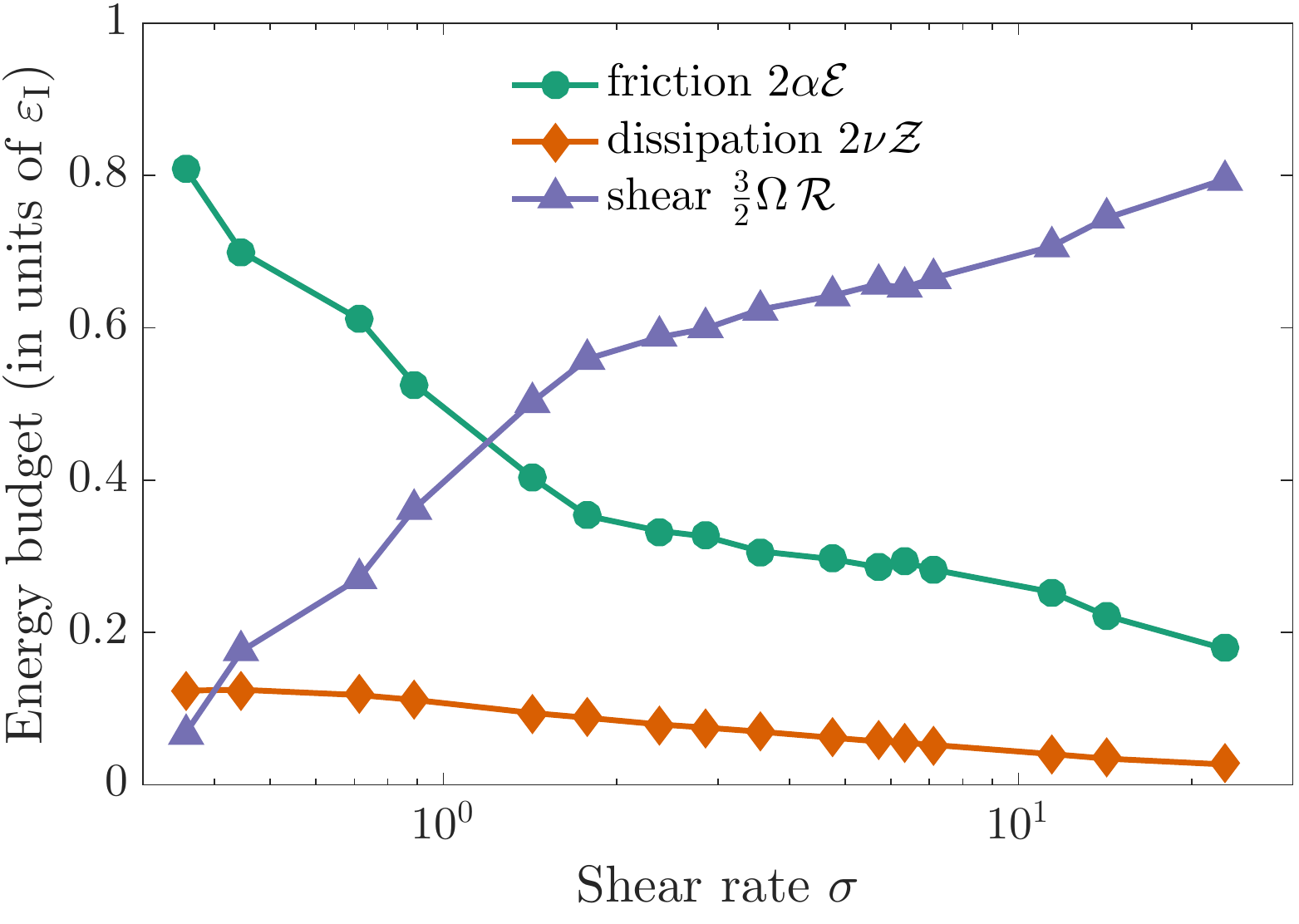}
    \caption{Contributions of friction, dissipation, and shear to the energy balance~\eqref{eq:energy_budget} as a function of the shear rate parameter~$\sigma$.}
    \label{fig:energy_budget}
\end{figure}
\begin{figure}[htpb]
    \centering
    \includegraphics[width=.45\textwidth]{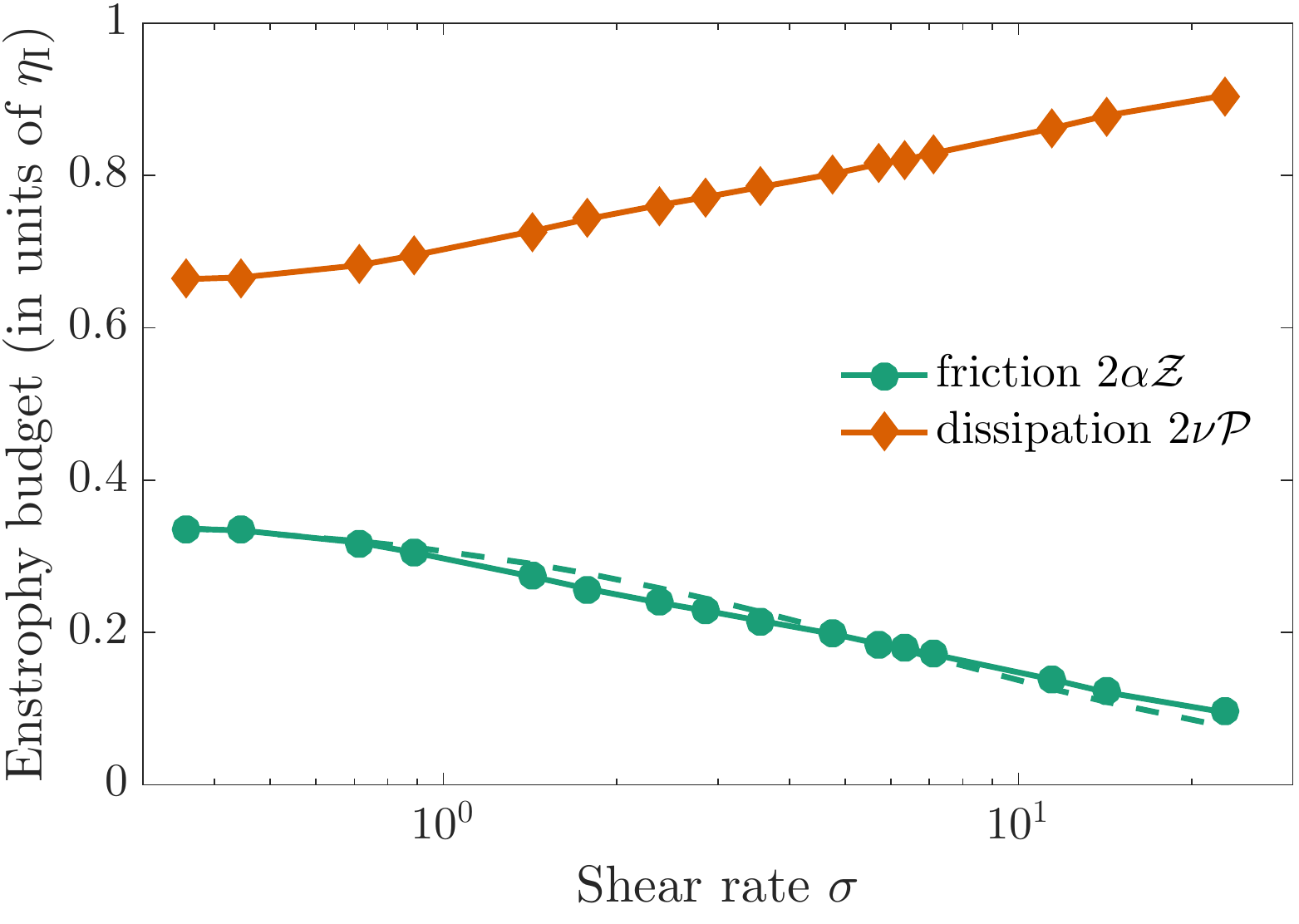}
    \caption{Friction and dissipation terms that enter the enstrophy balance~\eqref{eq:enstrophy_budget}, as a function of $\sigma$. The dashed line corresponds to a fit of the enstrophy of the form $\tau_{\rm f}^2\mathcal{Z}=(0.06-0.08\,\sigma)^{-1}$.}
    \label{fig:enstrophy_budget}
\end{figure}
We have performed several numerical experiments in which we varied the rotation rate $\Omega$ while keeping constant the forcing mechanisms, and thus the injection rates $\varepsilon_{\rm I}$ and $\eta_{\rm I}$. In all cases, transients are followed by the establishment of a statistical steady state associated with a direct cascade of enstrophy. These states are characterized by given values of $\mathcal{E}$, $\mathcal{Z}$, $\mathcal{P}$ and $\mathcal{R}$ that balance differently in the budgets. Figure~\ref{fig:energy_budget} shows the terms entering in the energy budget~(\ref{eq:energy_budget}) as a function of $\sigma=\tfrac{3}{2}\Omega\tau_{\rm f}$. It provides information about the large scales. The viscous dissipation, which is here proportional to $\mathcal{Z}$, decreases with shear. The contribution from shear through the Reynolds stress instead increases rapidly with $\sigma$, and the global balance then requires friction to decrease. The turbulent kinetic energy consequently decreases as a function of the shear rate parameter $\sigma$. The terms entering the enstrophy budget~(\ref{eq:enstrophy_budget}) are shown in Fig.~\ref{fig:enstrophy_budget}, giving this time information about small-scale turbulent fluctuations. Viscous dissipation increases with $\sigma$, because shear stretches vorticity, leading to larger values of the palinstrophy. The enstrophy balance implies that the friction consequently decreases with the rotation rate. The enstrophy $\mathcal{Z}$ also decreases and turbulent fluctuations weaken. 

\begin{figure}[htpb]
    \centering
    \includegraphics[width=.45\textwidth]{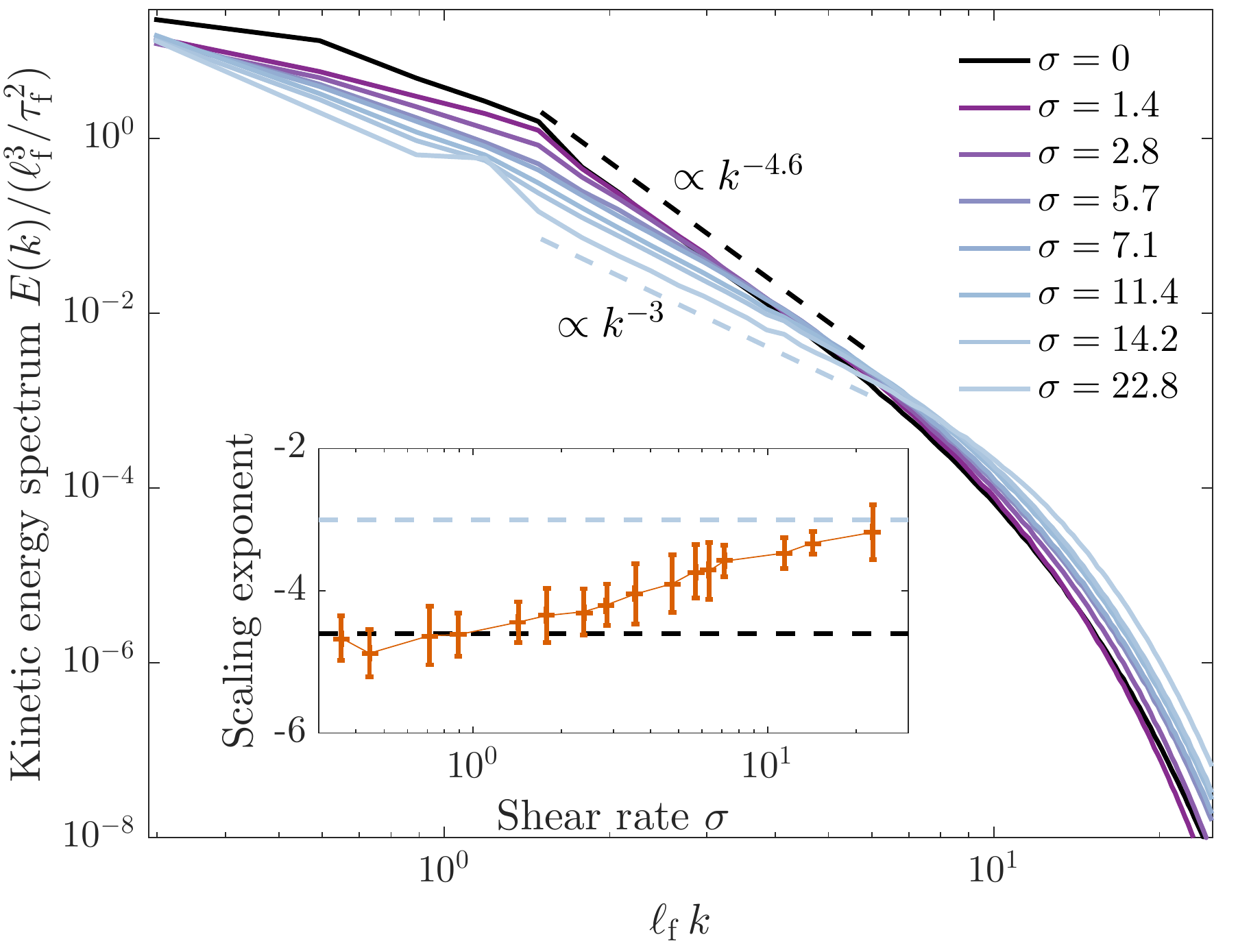}
    \caption{Time-averaged power spectra of the turbulent kinetic energy for various values of the shear parameter $\sigma$, as labelled. Inset: Measured scaling exponent that shows a continuous evolution from a value lesser than $-4.5$ to $-3$ when the shear rate increases.}
    \label{fig:spectrum}
\end{figure}
Figure \ref{fig:spectrum} reports measurements of the kinetic energy spectrum for the various values of the shear rate parameter $\sigma$. The direct cascade of enstrophy in two dimensions is in principle associated to a $k^{-3}$ scaling range for the spectrum (see, \textit{e.g.}, \cite{boffetta2012two}). This law pertains to the case $\alpha=0$. However, in the presence of friction, deviations from the -3 scaling have been found in numerics~\citep{BCMV02} and in experiments~\citep{BCEM05}, resulting in a steeper slope of the energy spectrum, with the exponent that decreases as a function of $\alpha$. Our numerics confirm such findings, with a scaling exponent $\approx-4.6$ for $\sigma=0$. As seen from the inset of Fig.~\ref{fig:spectrum}, the exponents increases with $\sigma$ and asymptotically reaches $-3$ for very large shear. This tendency can be explained from previous considerations on the energy and enstrophy budgets (Fig.~\ref{fig:energy_budget} and Fig.~\ref{fig:enstrophy_budget}) that show a reduction of the strength of friction when $\sigma$ increases, causing a lower effective value of $\alpha$. 

\begin{figure}[htpb]
    \centering
    \includegraphics[width=.45\textwidth]{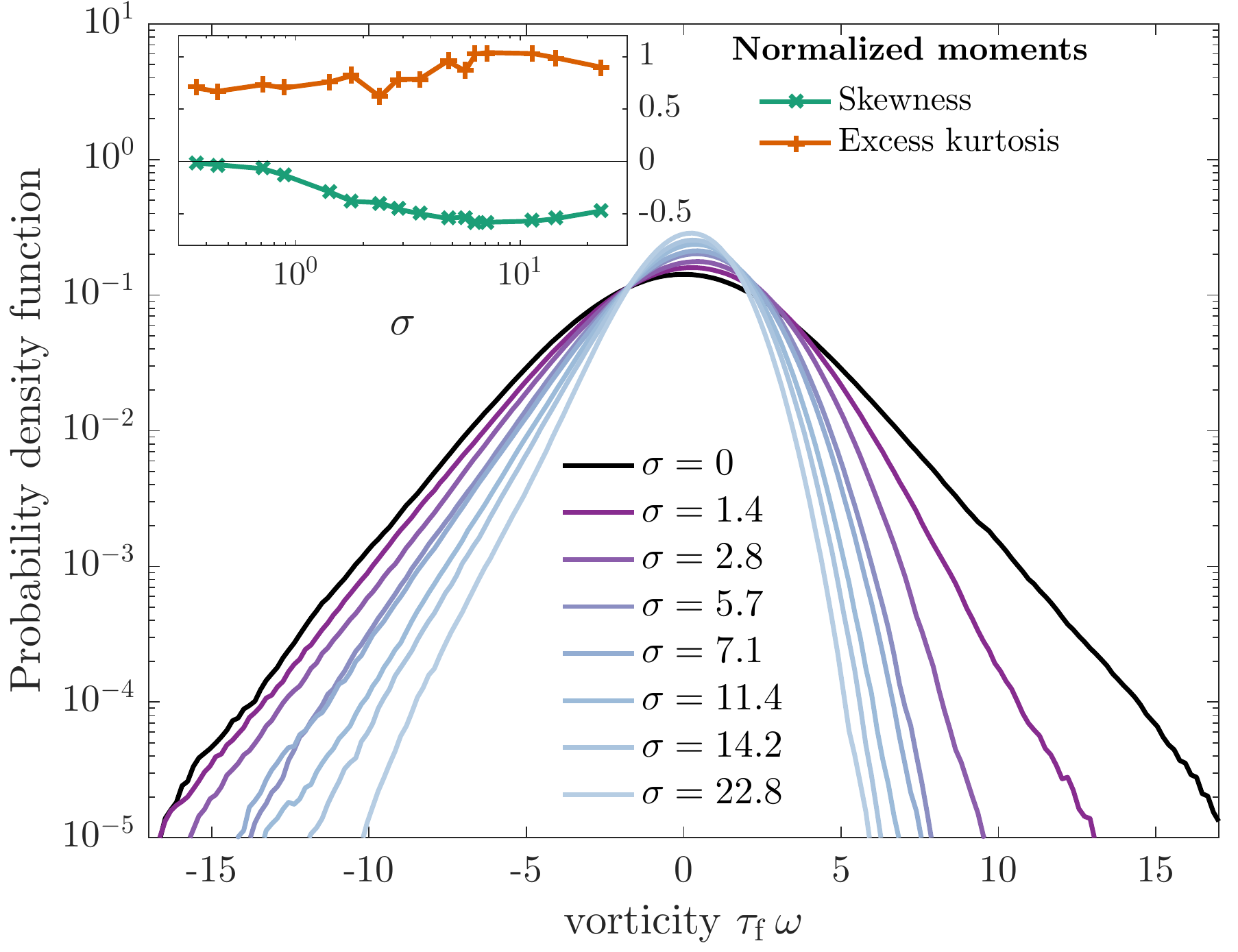}
    \caption{Probability density functions (PDF) of the vorticity for various values of the shear parameter $\sigma$. The inset shows the vorticity skewness and excess kurtosis as a function of the shear parameter.}
    \label{fig:pdf_vorticity}
\end{figure}
Figure \ref{fig:pdf_vorticity} shows the probability density functions (PDF) of the vorticity obtained when varying the shear parameter. The distributions become progressively asymmetrical when increasing $\sigma$, with broader negative tails, and reduced positive ones. This is clearly visible from the inset, which shows the skewness $\langle \omega^3\rangle/\langle \omega^2\rangle^{3/2}$ becoming more negative and the excess kurtosis $\langle \omega^4\rangle/\langle \omega^2\rangle^{2}-3$ increasing with $\sigma$. Shear tends to destroy cyclones (structures with a positive vorticity), favoring the survival of anticyclones. The shear indeed provides a constant negative input of  vorticity that entrains anticyclones and damps cyclones. This is at variance with simulations of purely rotating turbulence that favour the persistence of cyclonic vortexes (see, \textit{e.g.},~\citep{biferale2016coherent}). Keplerian shear overwhelms this effect. One notices in the figure's inset an unexpected change of tendency at the largest values of $\sigma$. We interpret this as geometrical effects that  appear at very fast rotations and which could require to consider boxes with an even larger aspect ratio than the one we assumed (4:1).

\begin{figure}[htpb]
    \centering
    \includegraphics[width=.45\textwidth]{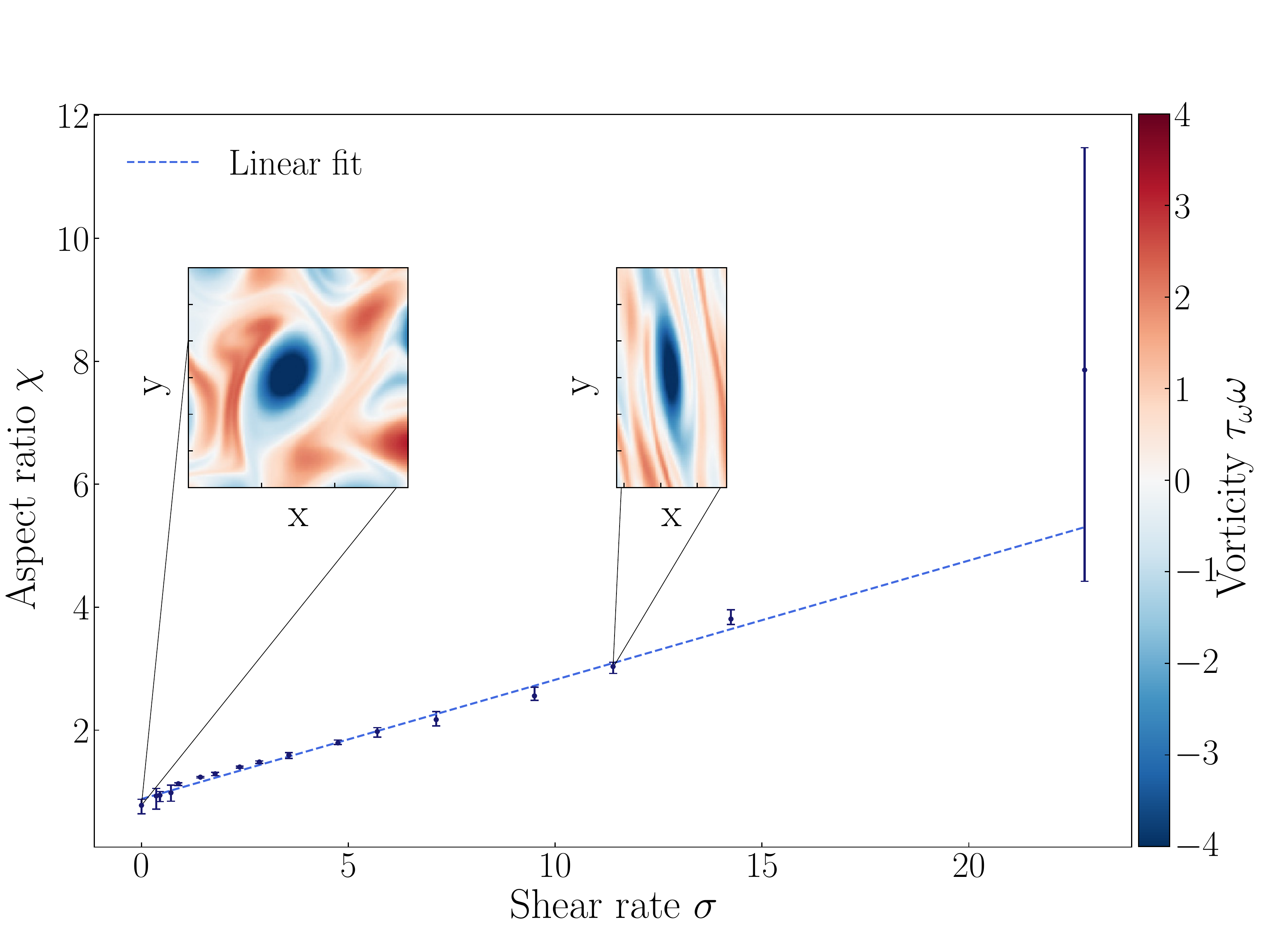}
    \caption{Aspect ratio $\chi$ of anticyclonic vortices as a function of the shear rate parameter $\sigma=\tfrac{3}{2}\Omega\tau_{\rm f}$. The solid line is a linear law fitting the data of the form $\chi = 1+0.19\,\sigma$. The two insets show snapshots of the vorticity perturbation $\omega$ (normalized by $\tau_{\omega}^{-1}$) for $\sigma=0$ and $\sigma=11.4$.}
    \label{fig:aspect_ratio}
\end{figure}

Above measurements suggest that the role played by anticyclonic vortices gets stronger when shear increases. As seen in the qualitative snapshots of Fig.~\ref{fig:flow}a, b, and c, the shape of such vortices varies with $\sigma$. It was found in~\cite{kida1981motion} that elliptical vortices are possible stable candidates to model such structures. They correspond to an elliptical patch of vorticity, whose major axis is aligned with shear (\textit{i.e.} with the $y$ direction), and with amplitude
\be
    \omega = -\frac{3}{2}\Omega \frac{\chi+1}{\chi\,(\chi-1)},
    \label{eq:kida_vorticity}
\ee
where $\chi\ge1$ denotes the aspect ratio between the major and minor axes. To estimate numerically this aspect ratio, we first identify elliptical points (where the gradient of the total velocity $\vec{v}$ has two complex-conjugate eigenvalues) corresponding to regions of the flow with a strong-enough rotation. We then compute the vorticity at such locations and invert \eqref{eq:kida_vorticity} to express a local value of the aspect ratio $\chi$, which is then averaged in both space and time. Figure \ref{fig:aspect_ratio} shows the dependence of this quantity on the shear rate parameter $\sigma = \tfrac{3}{2}\Omega\tau_{\rm f}$. These results indeed quantify the fact that anticyclones get stretched with increasing shear. Error bars are estimated from partial time averages over $1/5$ of the total duration of the simulations. Measurements are very well approximated by a linear fit. When fitting the data, we did not consider the largest value of the rotation rate ($\sigma= 22.8$), for which the above procedure leads to a very noisy estimate. Two snapshots of the vorticity are shown to illustrate the typical shape of anticyclones at $\sigma=0$ and $\sigma= 11.4$.

\section{Inertial particles in a Keplerian flow}
\label{section:particles}
We consider tiny dust particles whose dynamics in position-velocity $(\vec{R}_{\rm p},\vec{V}_{\rm p})$ phase space follows
\begin{eqnarray}
    \frac{d\vec{R}_{\rm p}}{dt}&=&\vec{V}_{\rm p}, \qquad \vec{R}_{\rm p} = (x_{\rm p},y_{\rm p})^\mathsf{T},\\
    \frac{d\vec{V}_{\rm p}}{dt}&=&-\frac{1}{\tau_{\rm p}}\left[\vec{V}_{\rm p}\!-\!\vec{v}(\vec{R}_{\rm p},t)\right]\!-\!2\vec{\Omega}\!\times\! \vec{V}_{\rm p}\!+\!3\Omega^2x_{\rm p}\vec{e}_x.
\label{part_eq}
\end{eqnarray}
The three forces appearing in the right-hand side of \eqref{part_eq} are the linear drag with the gas flow, the Coriolis force, and the combination of gravity and centrifugal forces, respectively.

The drag is proportional to the difference between the particle velocity and the total gas flow. It involves the particle response time $\tau_{\rm p}$, whose expression depends on the considered regime.  If the particle radius $a$ is large-enough compared to the gas mean-free path, their drag follows Stokes regime and $\tau_{\rm p} = (2/9) \rho_{\rm p}a^2/(\rho_{\rm g}\nu)$, where $\rho_{\rm p}$ denotes the material mass density of dust grains. Smaller radii follow the Epstein regime where $\tau_{\rm p} = (8/\pi)^{1/2}\rho_{\rm p}a/(\rho_{\rm g}c_{\rm s})$, with $c_{\rm s}$ denoting the speed of sound.  Neglecting both particle-particle interactions and their feedback onto the gas means that we do not need to specify the particle size, making our results relevant to both regimes.  The dependence upon the particle response time is usually expressed in terms of a Stokes number obtained by non-dimensionalizing $\tau_{\rm p}$ with a characteristic time of the flow. The usual choice in astrophysics is to introduce $St_{\Omega} = \tau_{\rm p}\Omega$. This Stokes number is mostly relevant at large scales, because it compares the particle response time to the rotation rate, which corresponds to the largest timescale of the flow. Another choice consists in defining $St_{\rm f} = \tau_{\rm p}/\tau_{\rm f} = \eta_{\rm I}^{1/3}\tau_{\rm p}$. As explained in the previous subsection, the forcing time $\tau_{\rm f}$ is actually the only timescale that is fixed in our simulation protocol. Finally, we will mostly rely on the small-scale Stokes number $St_{\omega} = \tau_{\rm p}/\tau_{\omega} = \langle\omega^2\rangle^{1/2}\tau_{\rm p}$, which involves the typical amplitude of turbulent fluctuations. Remember that in our protocol, the enstrophy decreases with $\Omega$, so that for a fixed response time, $St_\omega$ decreases with shear.

As assumed in the previous section, the gas moves in average with the Keplerian speed. Particles which follow \eqref{part_eq} do so, and there is hence no mean drag that would lead to a radial drift. Indeed, if we assume that there are no turbulent fluctuations, the gas velocity simply reads $\vec{v}(\vec{R}_{\rm p},t) = -\tfrac{3}{2}\Omega\,x_{\rm p}\vec{e}_y$ and one can easily see that particles moving with the Keplerian shear define stationary solutions to the dynamics. The presence of turbulent fluctuations make particles deviate from the gas streamlines and possibly concentrate. A particular role is played by rotating coherent structures. In usual hydrodynamic situations, vortices act as small centrifuges and normally eject heavy inertial particles. In the present case, the Coriolis force can actually counteract this effect and can sometimes even prevail over the local drag responsible for ejection. This mechanism can be qualitatively understood in terms of Maxey's approximation for the particle velocity~\cite{maxey1987gravitational}. At small values of $\tau_{\rm p}$, and in the presence of the Coriolis force, the particle velocity can be approximated as $\vec{V}_{\rm p} \approx \vec{v} - \tau_{\rm p}({\rm d}\vec{v}/{\rm d}t + 2 \vec{\Omega}\!\times\!\vec{v})$. When $\vec{\Omega}=0$, the term $\propto\tau_{\rm p}$ is responsible for driving particles out of vortices. It is however well-known that in the presence of rotation, this term can actually change sign if the Coriolis force overwhelms the fluid acceleration. Particles are then attracted towards the center of the vortices. Such an effect requires that ${\rm d}\vec{v}/{\rm d}t$ and $\vec{\Omega}\!\times\!\vec{v}$ have opposite directions, so that it only occurs in anticyclonic vortices.

Previous qualitative argument suggests that particle concentration in anticyclones requires a strong-enough rotation. However, it does not tell anything about dependence on their Stokes number. To get further insight, and building on the results of the previous subsection, we next examine how particles cluster in an elliptical patch of vorticity. We follow again the normal form proposed in \cite{kida1981motion} and assume that, locally inside a vortex centered at the origin, the fluid velocity can be approximated as
\begin{equation}
    \vec{v} = \mathbb{A}\begin{bmatrix}x\\ y \end{bmatrix}, \mbox{ with } \mathbb{A} = \frac{3}{2}\frac{\Omega}{\chi-1} \begin{bmatrix} 0 &  \chi^{-1} \\ -\chi & 0\end{bmatrix}.
\end{equation}
As previously noticed, the aspect ratio $\chi$ itself depends on the rotation rate $\Omega$. We hereafter use the linear fit proposed in Fig.~\ref{fig:aspect_ratio}. The next step consists in analyzing the stability of the linear system
\begin{equation}
\frac{\rm d}{{\rm d}t} \begin{bmatrix}
    \vec{{R}_{\rm p}}\\
    \vec{{V}_{\rm p}}
\end{bmatrix}
    = \begin{bmatrix}
    \mathbb{0} & \mathbb{1} \\
    \tfrac{1}{\tau_{\rm p}} \mathbb{A} +\!3\Omega^2 \mathbb{e}_{xx} & -\tfrac{1}{\tau_{\rm p}}\mathbb{1} -2\Omega\mathbb{J}
    \end{bmatrix}
    \begin{bmatrix}
    \vec{R_{\rm p}}\\
    \vec{V_{\rm p}}
\end{bmatrix},
\label{eq:tangent_dyn}
\end{equation}
where $\mathbb{e}_{xx} = [1\ 0; 0\ 0]$ and $\mathbb{J} = [0\ -1; 1\ 0]$. The vortex center is stable only if all the eigenvalues of the tangent matrix appearing in \eqref{eq:tangent_dyn} have a negative real part. 

\begin{figure}[htpb]
    \centering
    \includegraphics[width=.45\textwidth]{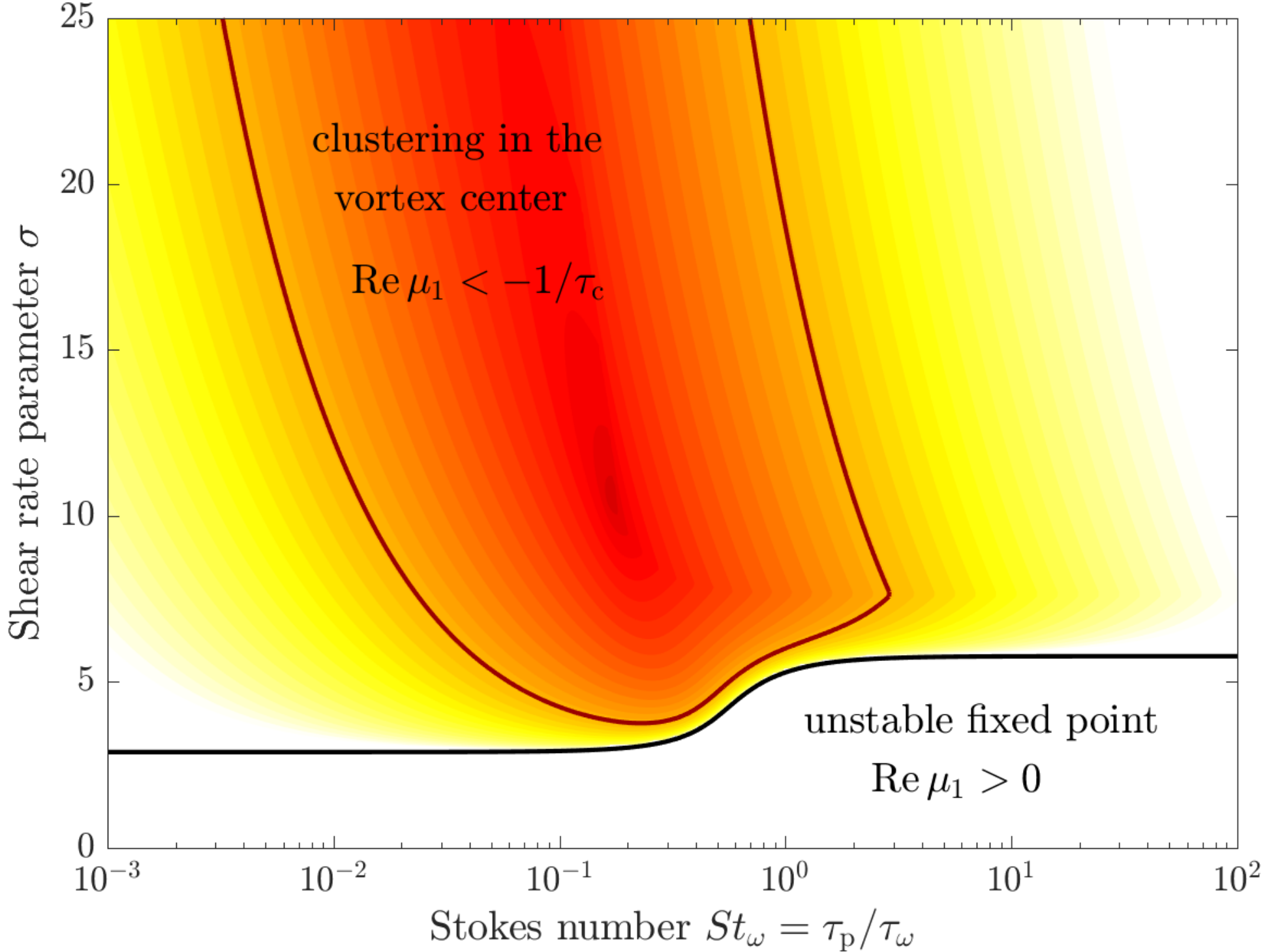}
    \caption{Linear analysis of particles dynamics in a stationary vortex patch given by \eqref{eq:tangent_dyn}: Contour plots of the eigenvalue $\mu_1$ with largest real part in the Stokes number / shear rate plane. The black solid line corresponds to ${\rm Re}\,\mu_1=0$, while the red one delimit a region where $-{\rm Re}\,\mu_1$ is larger than a given threshold (see text).}
    \label{fig:kida_analysis}
\end{figure}

As shown in Fig.~\ref{fig:kida_analysis}, this fixed point is always unstable at the smallest values of  $\Omega$ (below the black solid line), confirming the idea that vortices eject particles.
Above this line, vortex centers are found to always be stable fixed points of the particle dynamics. The convergence rate towards the origin however depends on both the particle response time and the rotation rate. It can be estimated from the real part of the less contracting eigenvalue $\mu_1$, which is shown as a colored background in Fig.~\ref{fig:kida_analysis}. The parameters used in this figure were chosen to match our direct numerical simulations. For instance, we have used the fit of Fig.~\ref{fig:enstrophy_budget} to express $\tau_{\omega}$ as a function of $\Omega$. The value of the contraction rate does not give much information, unless it is compared to another characteristic time of the dynamics. Indeed, if we assume that the considered vortex has only a finite lifetime $\tau_{\rm c}$, convergence to the vortex center can only occur if the contraction rate is large enough, namely $-{\rm Re}\,\mu_1 > 1/\tau_c$ (illustrated as a red solid line in the figure). Hence, for a fixed value of the shear parameter, only an interval of Stokes numbers can actually lead to strong clustering. At small Stokes numbers, inertia is too weak and particles are migrating too slowly toward the vortex center, namely on timescales that exceed the lifetimes of structures. At large values, particles spiral many times around the origin, making again convergence too slow. As we will now see, such considerations are of relevance to describe clustering in actual turbulent flows.

\section{Ejection vs.\ concentration in vortices}
\label{section:fractvsstrong}
We now turn to detect and characterize in our simulations the effects of coherent vortices on the particle distribution. As reported in Sec.~\ref{section:shearingbox}, we have performed 18 different sets of simulations where we varied the rotation rate $\Omega$, while keeping constant the forcing characteristics and all other flow parameters. In each simulation, 24 different families of Lagrangian particles with different response times were integrated. They are passive (no feedback onto the flow) and do not interact with each other. 

\begin{figure}[htpb]
    \centering
    \hspace{-0.5cm}
    \includegraphics[width=.5\textwidth]{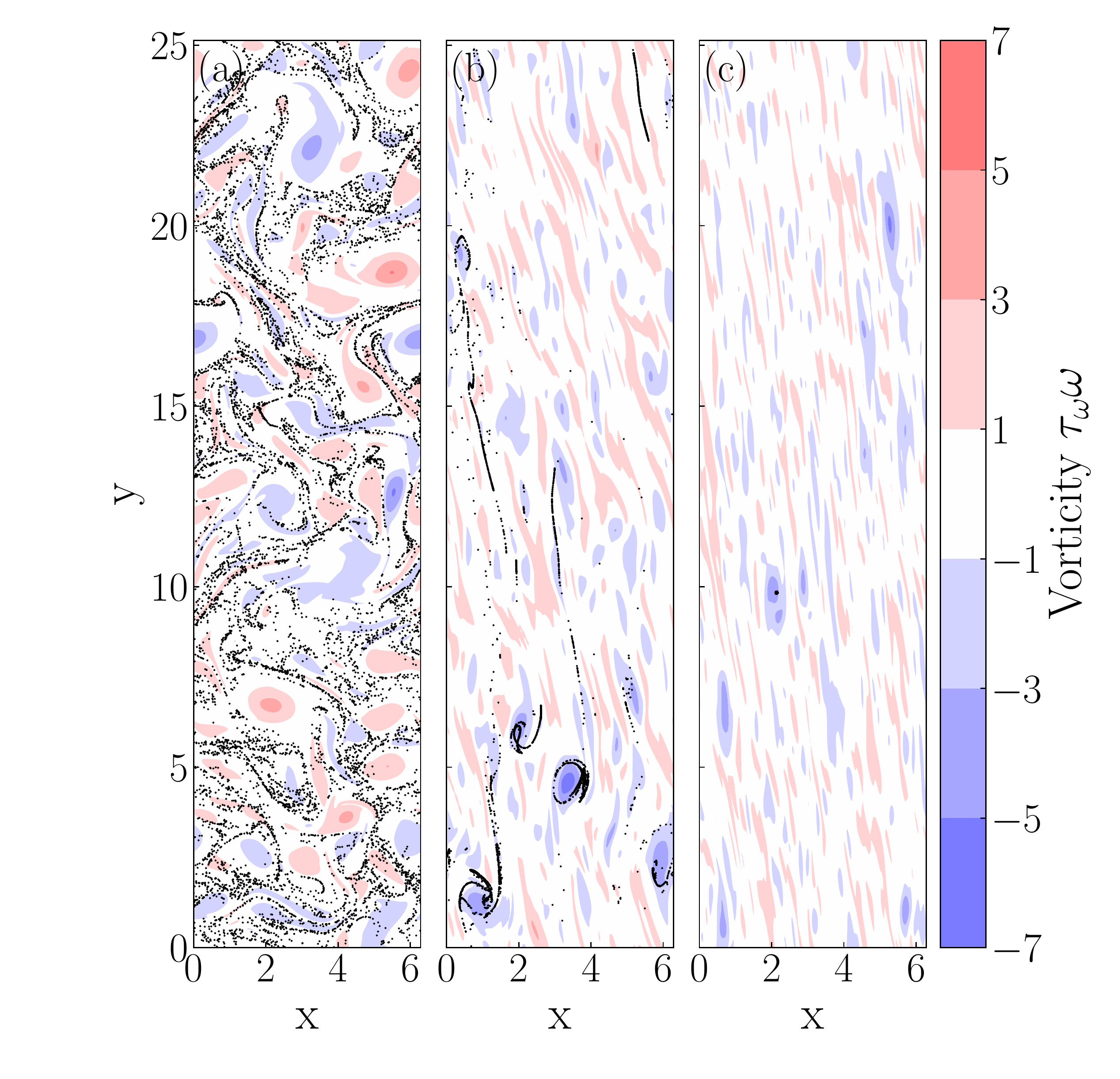}
    \caption{Vorticity $\omega$, normalized by $\tau_{\omega}^{-1}$, and dust particle position $\vec{R}_{\rm p}$ (black dots) for $St_{\omega}\approx 0.35$ and (a) $\sigma$=0, (b) $\sigma$=5.7, and (c) $\sigma$=14.2.}
    \label{fig:movie}
\end{figure}

Figure~\ref{fig:movie}a, b, and c show the instantaneous spatial positions of particles in the statistically stationary regimes, for a fixed value of the Stokes number $St_{\omega}$ and increasing the shear rate. 
The color code is for the fluid vorticity $\omega$ (normalised by $\tau_{\omega}^{-1}$), while the positions of dust particles are shown as black dots. 
In the absence of shear and rotation (Fig. \ref{fig:movie}a), particle positions are clearly expelled from vortices, independently of their sign, and concentrate along stripes in between them. At the larger, intermediate values of rotation  ($\sigma=5.7$, Fig.~\ref{fig:movie}b), particles are still avoiding vortex cores but distribute this time in spirals around the anticyclones (with a negative vorticity, in blue). Finally, for the largest rotation rate ($\sigma=14.2$, Fig. \ref{fig:movie}c), all particles have converged in a unique point cluster located in the core of an anticyclone. Increasing the rotation rate has a drastic impact on whether coherent vortices act as centrifuges or give birth to point clusters.

Such observations can be quantified by measuring fluid velocity characteristics along particle paths. The persistent structures of the flow are often linked to its local structure through the Okubo--Weiss parameter, $OW$ (see, \textit{e.g.},~\cite{perlekar2011persistence}). It is defined from the (possibly imaginary) eigenvalues $\pm\beta$ of the strain tensor $\textrm{A}_{ij}=\partial_j\rm{v}_i$, namely
\begin{equation}
    OW=-\beta^2=-(\partial_x\textrm{v}_x)^2-\partial_x\textrm{v}_y\partial_y\textrm{v}_x
\end{equation}
This parameter can change sign and therefore divides the flow between $OW<0$ (straining, hyperbolic regions, where the two opposite eigenvalues are real) and $OW>0$ (rotating, elliptic regions, where the strain has purely imaginary eigenvalues).

\begin{figure}[htpb]
    \centering
    \includegraphics[width=.35\textwidth]{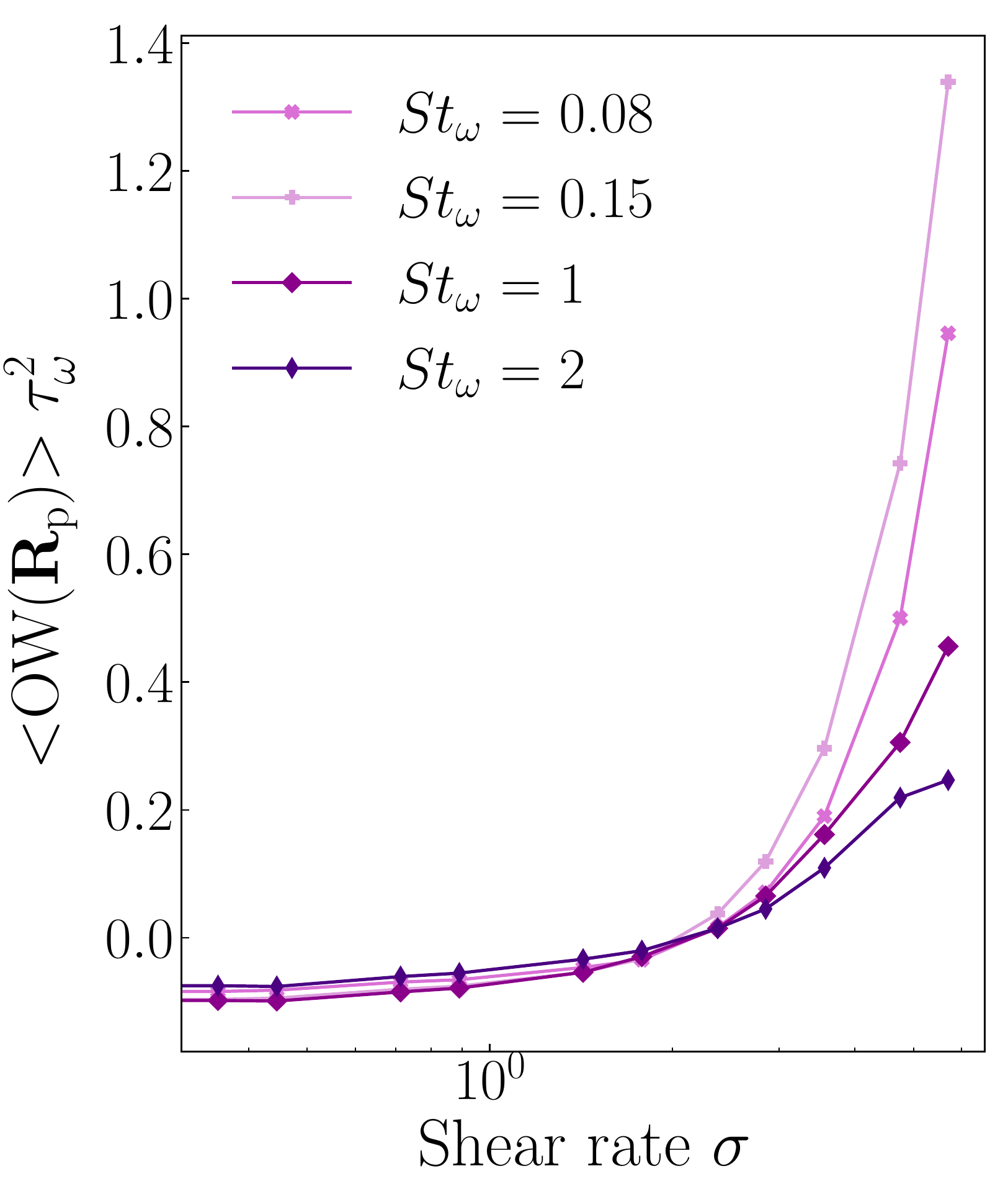}
    \caption{Average Okubo--Weiss parameter at particle position $OW(\vec{R}_{\rm p})$ as a function of the shear rate parameter $\sigma$ for various values of the Stokes number $St_{\omega}$.}
    \label{fig:OW}
\end{figure}

The Okubo--Weiss parameter has been interpolated at particle positions. Figure~\ref{fig:OW} shows its average value $\langle OW(\vec{R}_{\rm p}(t),t)\rangle$ over all particles and times, as a function of the shear rate and for various values of the Stokes number.
Independently of $St_{\omega}$, this average is negative at small values of $\sigma$, meaning that inertial particles spend most of the time in the straining, hyperbolic regions of the flow. However, the average Okubo--Weiss parameter becomes positive for $\sigma\gtrsim 2$ and this threshold seems to depend only weakly on the Stokes number. This is consistent with the linear analysis of Kida's vortices from the previous section, which predicts that anticyclones become stable for large-enough shears.  At even larger rotation rates, the  average Okubo--Weiss parameter shows a fast increase, especially at small values of $St_{\omega}$. Dust particles are therefore living mostly inside rotating, elliptic regions of the flow. 

\begin{figure}[htpb]
    \centering
    \includegraphics[width=0.45\textwidth]{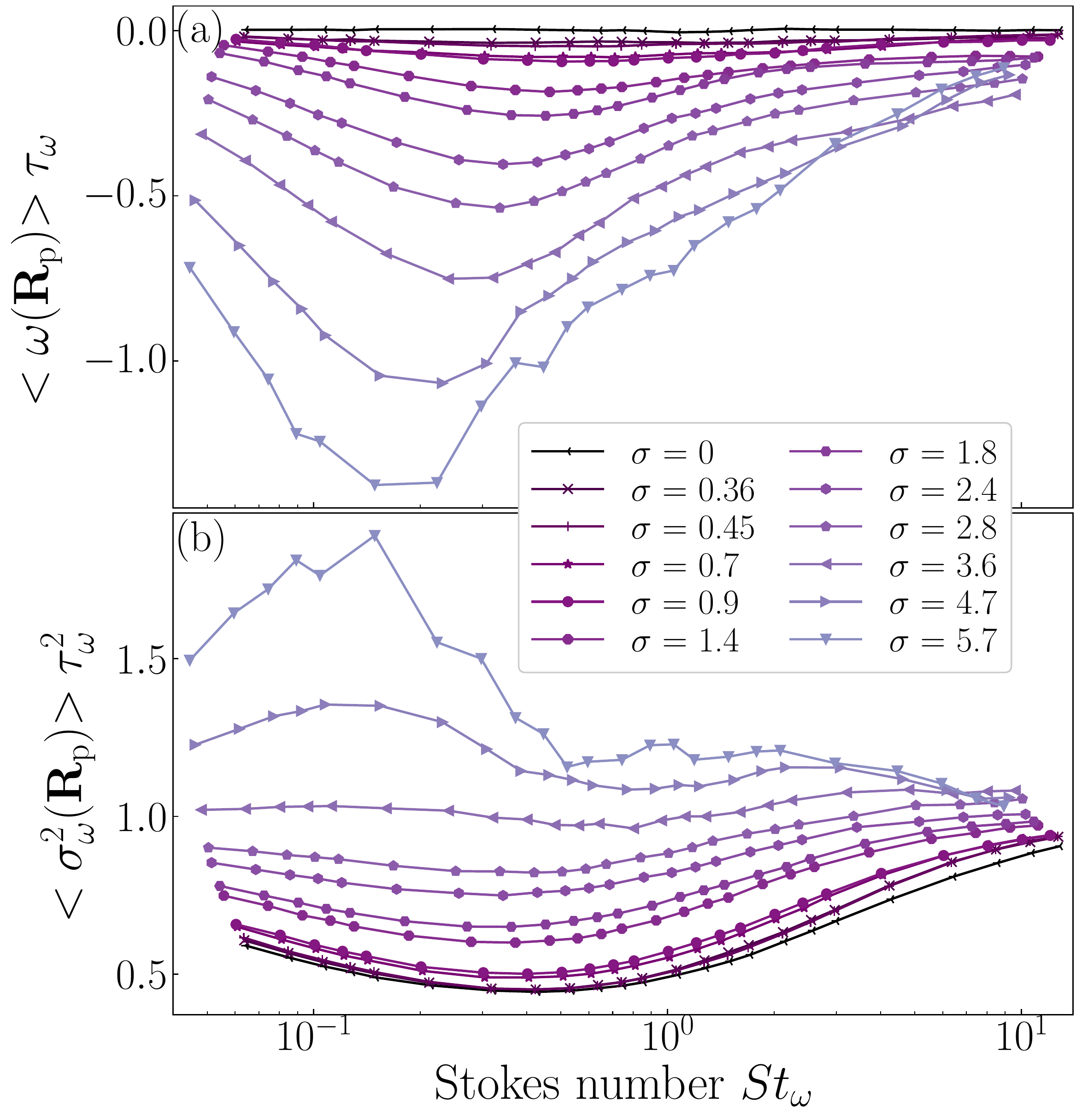}
    \caption{Mean (a) and variance (b) of vorticity $\omega(\vec{R}_{\rm p}(t),t)$ at particle position, as a function of the Stokes number $St_{\omega}$, for various values of the shear rate parameter $\sigma=\tfrac{3}{2}\Omega\tau_{\rm f}$. }
    \label{fig:vorticity_moms}
\end{figure}

A complementary information on the nature of these eddies can be obtained from the statistics of the fluid vorticity at particle position $\omega(\vec{R}_{\rm p}(t),t)$. The mean and variance over all particles and times, are reported in Fig.~\ref{fig:vorticity_moms}a and b, as a function of the Stokes number and for various rotation rates. In the absence of rotation ($\sigma=0$), the average $\langle\omega(\vec{R}_{\rm p}(t),t)\rangle$ almost vanishes for all Stokes numbers, giving an estimate of the statistical convergence of our data. In the presence of rotation, $\langle\omega(\vec{R}_{\rm p}(t),t)\rangle$ is always negative, and this bias increases with $\sigma$. Significantly negative average vorticities indicate that the particles spend most of the time in the anticyclones. This happens for fast-enough rotations and intermediate values of $St_{\omega}$. It is for such parameter values that anticyclonic vortices are expected to concentrate particle trajectories, as anticipated in the previous section from the analysis of Kida vortices.
The variance of vorticity at particle position is shown in Fig.~\ref{fig:vorticity_moms}b. At small rotation rates, particles sample flow regions with lesser vorticity fluctuations. At larger values of $\sigma$, they tend to concentrate in regions where vorticity has strong fluctuations. This effect again peaks at intermediate values of the Stokes numbers. 

\begin{figure}[htpb]
    \centering
    \includegraphics[width=0.45\textwidth]{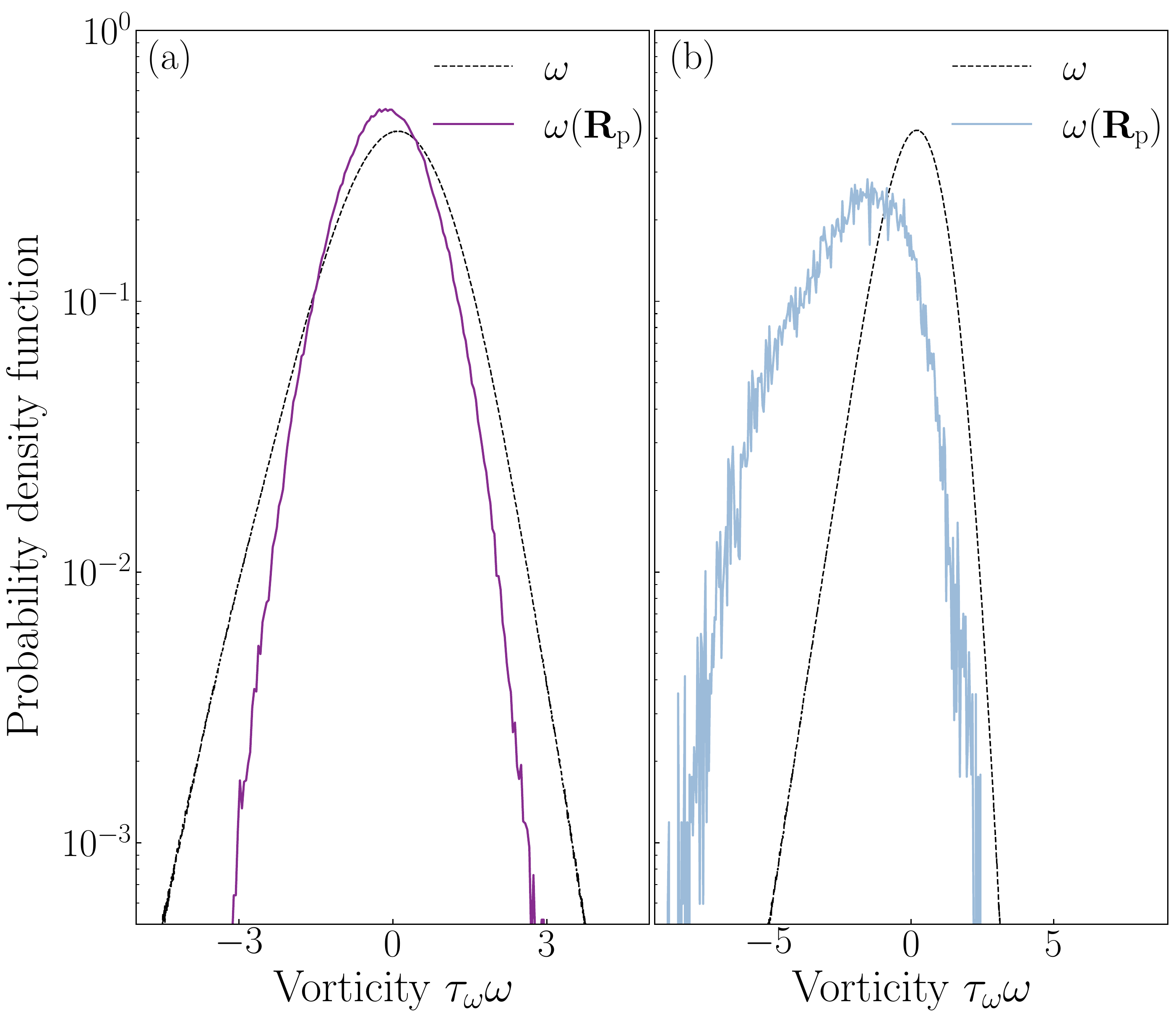}
    \caption{PDFs of the Eulerian vorticity $\omega$ and of the Lagrangian one $\omega(\vec{R}_{\rm p}(t),t)$ for $St_{\omega}\approx0.19$ and (a) $\sigma=1.4$, (b) $\sigma=11.4$.}
    \label{fig:vorticity_pdf}
\end{figure}

Finally, Figure~\ref{fig:vorticity_pdf}a and b show the probability density functions of the vorticity, for two different values of $\sigma$. The PDFs are computed both for the fluid and at the particle position.  At low rotation ($\sigma=1.4$, Fig.~\ref{fig:vorticity_pdf}a), both distributions are almost symmetric. Lagrangian statistics display narrower tails than the Eulerian ones, indicating that particle are preferentially sampling the less active regions of the flow. At the larger rotation ($\sigma=11.4$, Fig.~\ref{fig:vorticity_pdf}b), both distributions are skewed toward negative values. However, Lagrangian statistics display a broader negative tail than the Eulerian ones, indicating this time that particles are spending more time in the most active regions of the flow. Skewness toward negative values moreover tells that particles concentrate in anticyclonic vortices, as it could already be seen qualitatively in Fig.~\ref{fig:movie}c.

Despite its apparently noisy aspect, the Lagrangian vorticity distribution shown for $St_{\omega}=0.19$ and $\sigma=11.4$ in Fig.~\ref{fig:vorticity_pdf}b has actually required a tremendous amount of statistics. For these specific values of the parameters, we have actually performed, in addition to time averages, ensemble average over 20 different realisations of the dynamics. The convergence of all particles in the core of a single, long-living, structure makes average with respect to particles meaningless. For that reason, the results on the mean and variance of the Okubo--Weiss parameter and of the vorticity at particle position (in Fig.~\ref{fig:OW} and Fig.~\ref{fig:vorticity_moms}a and b) have been displayed only up to $\sigma\approx5$. Simulations at larger values of the rotation rate led to much noisier results. As we will see in the next section, the concentration of particles in point clusters can actually be quantified in terms of other observables whose statistics are not affected by such pitfalls.

\section{Fractal mass distribution}
\label{section:fract}
Results from the previous section indicate that for large-enough rotation rates and intermediate values of their Stokes number, particles preferentially sample the rotating, coherent structures of the flow. This is in line with the linear stability analysis of model vortices that is made in Sec.~\ref{section:particles}. This local analysis, which of course does not straightforwardly apply to fluctuating, time-varying flows, can still be generalized by considering the tangent system to particles' dynamics. The four-dimensional phase-space separation $\boldsymbol{\Delta}_t = (\delta\vec{R}_{\rm p}(t), \delta\vec{V}_{\rm p}(t))^{\sf T}$ between two infinitesimally close particle trajectories follows 
\begin{equation}
    \frac{\rm d}{{\rm d}t} \boldsymbol{\Delta}_t =\mathbb{M}_{t}\,\boldsymbol{\Delta}_t, 
    \label{eq:evol_sep}
\end{equation}
\begin{equation}
    \mbox{with}\quad
\mathbb{M}_{t}   
    = \begin{bmatrix}
    \mathbb{0} & \mathbb{1} \\
    \tfrac{1}{\tau_{\rm p}} \mathbb{A}(t) +\!3\Omega^2 \mathbb{e}_{xx} & -\tfrac{1}{\tau_{\rm p}}\mathbb{1} -2\Omega\mathbb{J}
    \end{bmatrix},
\end{equation}
where $\mathbb{e}_{xx} = [1\ 0; 0\ 0]$, $\mathbb{J} = [0\ -1; 1\ 0]$, and $\mathbb{A}$ is the gas velocity gradient tensor computed along a reference particle trajectory, namely $\textrm{A}_{ij}(t) = \partial_j\textrm{v}_i (\vec{R}_{\rm p}(t),t)$. The linear evolution~\ref{eq:evol_sep} integrates to
\begin{equation}
    \boldsymbol{\Delta}_t=\mathcal{J}_{t}\,\boldsymbol{\Delta}_{0}, \quad \mbox{with} \quad \mathcal{J}_{t}=\mathcal{T}\!\exp \int_{0}^{t}\textrm{d}\tau\,\mathbb{M}_{\tau}.
    \label{eq:tangent_syst}
\end{equation}
$\mathcal{J}_{t}$ is the Green operator associated to the linearised dynamics, with $\mathcal{T}\!\exp$ the time-ordered exponential. The matrix $\mathcal{J}^{\sf T}_{t}\mathcal{J}_{t}$ has positive eigenvalues that can be written in the form ${\rm e}^{2\mu_j(t)t}$ with $j=1..4$. The exponents $\mu_j$ are called the stretching rates. The Lyapunov exponents are then obtained from the ordered long-time behaviour of the stretching rates: $\lambda_j=\lim_{t\rightarrow\infty}\mu_j(t)$, with $\lambda_1\ge\lambda_2\ge\lambda_3\ge\lambda_4$.

The largest Lyapunov exponent $\lambda_1$ measures the growth rate of  distances between infinitesimally close trajectories. When $\lambda_1>0$ the dynamics is chaotic, while for $\lambda_1<0$ phase-space volumes contract and all particle trajectories tend toward each other. Higher-order Lyapunov exponents generalise these notions to infinitesimal phase-space surfaces, volumes, etc. that can either expand or contract. For dissipative dynamics, they can be used to define the Lyapunov dimension~\cite{KY78} as
\begin{equation}
    D_{\rm L}=J-\frac{\lambda_1+\cdots+\lambda_J}{\lambda_{J+1}}
\end{equation}
where the integer $J$ is such as $\lambda_1+\cdots+\lambda_J\geq0$ and $\lambda_1+\cdots+\lambda_{J+1}<0$. This quantity estimates the dimension of the attractor in phase space and has been already successfully used to quantify fractal concentrations of inertial particles~\cite{bec2003fractal,boffetta2004large,gustavsson2011ergodic}.  Depending on $D_{\rm L}$, three different regimes for the distribution of particles can be defined. If $D_{\rm L}>d$, the attractor has a dimension larger than physical space. As particle positions are obtained by projecting one set onto the other, particles fill in that case the whole space. Instead, if $D_{\rm L}<d$ the particles form a fractal set in position-velocity that keeps its fractal structure upon projection, leading to fractal concentrations. Finally, when $D_{\rm L}=0$ particles form point clusters in both phase space and positions. Notice that $D_{\rm L}$ cannot take values between 0 and 1. Indeed, if $\lambda_1>0$, the attractor has to be at least a line. On the contrary, if $\lambda_1<0$ the dimension of the attractor has to be $0$.

\begin{figure}[htpb]
    \centering
    \includegraphics[width=\columnwidth]{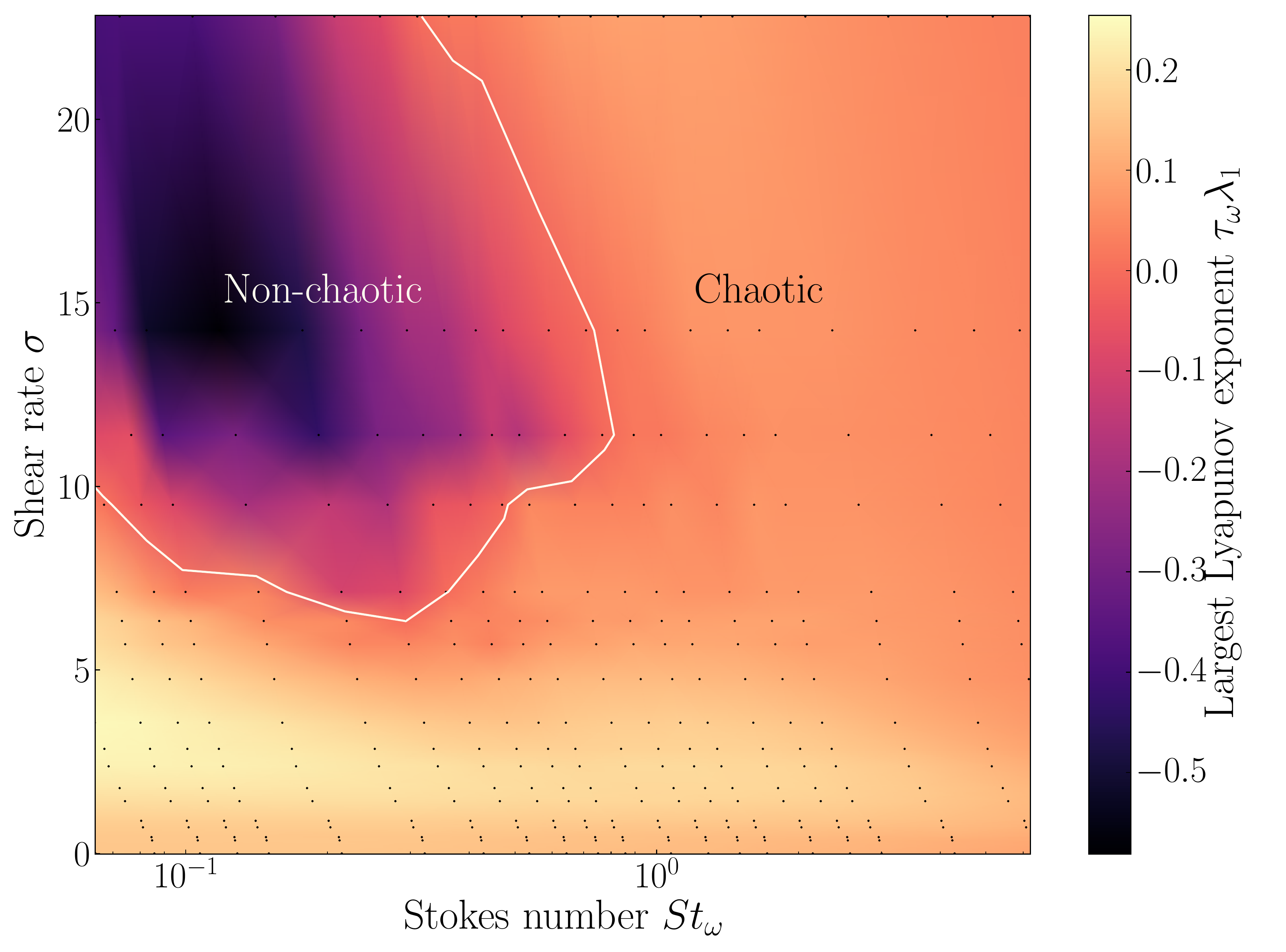}
    \caption{Phase diagrams of the shear rate parameter $\sigma$ vs the Stokes number $St_{\omega}$. The color code is for the largest Lyapunov exponent $\lambda_{1}$ (normalized by $\tau_{\omega}^{-1}$). Each black dot stands for the parameters that are actually simulated.}
    \label{fig:lyapunov_exp}
\end{figure}
We have evaluated the Lyapunov exponents in our numerical simulations by integrating the tangent dynamics~(\ref{eq:evol_sep}) along a large set of particle trajectories. By computing the stretching rates $\mu_j(t)$ with the standard method of~\citep{BGGS80}, we obtain the Lyapunov exponents as their average over time and particles. Figure~\ref{fig:lyapunov_exp} represents how the largest Lyapunov exponent $\lambda_1$ (expressed in units of $\tau_\omega^{-1}$) depends on both the shear parameter $\sigma$ and the particle Stokes number $St_{\omega}$. One observes that at large-enough values of the particle response time, $\tau_\omega\lambda_1$ decreases with the rotation rate. This depletion of chaoticity mainly comes from the fact that in this limit, turbulent fluctuations get depleted by shear, as already observed in Sec.~\ref{section:shearingbox}. Another noticeable behaviour can be observed at sufficiently large values of $\sigma$ and small to intermediate $St_{\omega}$ (top-left region). We find there that $\lambda_1<0$ and particle dynamics looses its chaoticity.

\begin{figure}[htpb]
    \centering
    \includegraphics[width=\columnwidth]{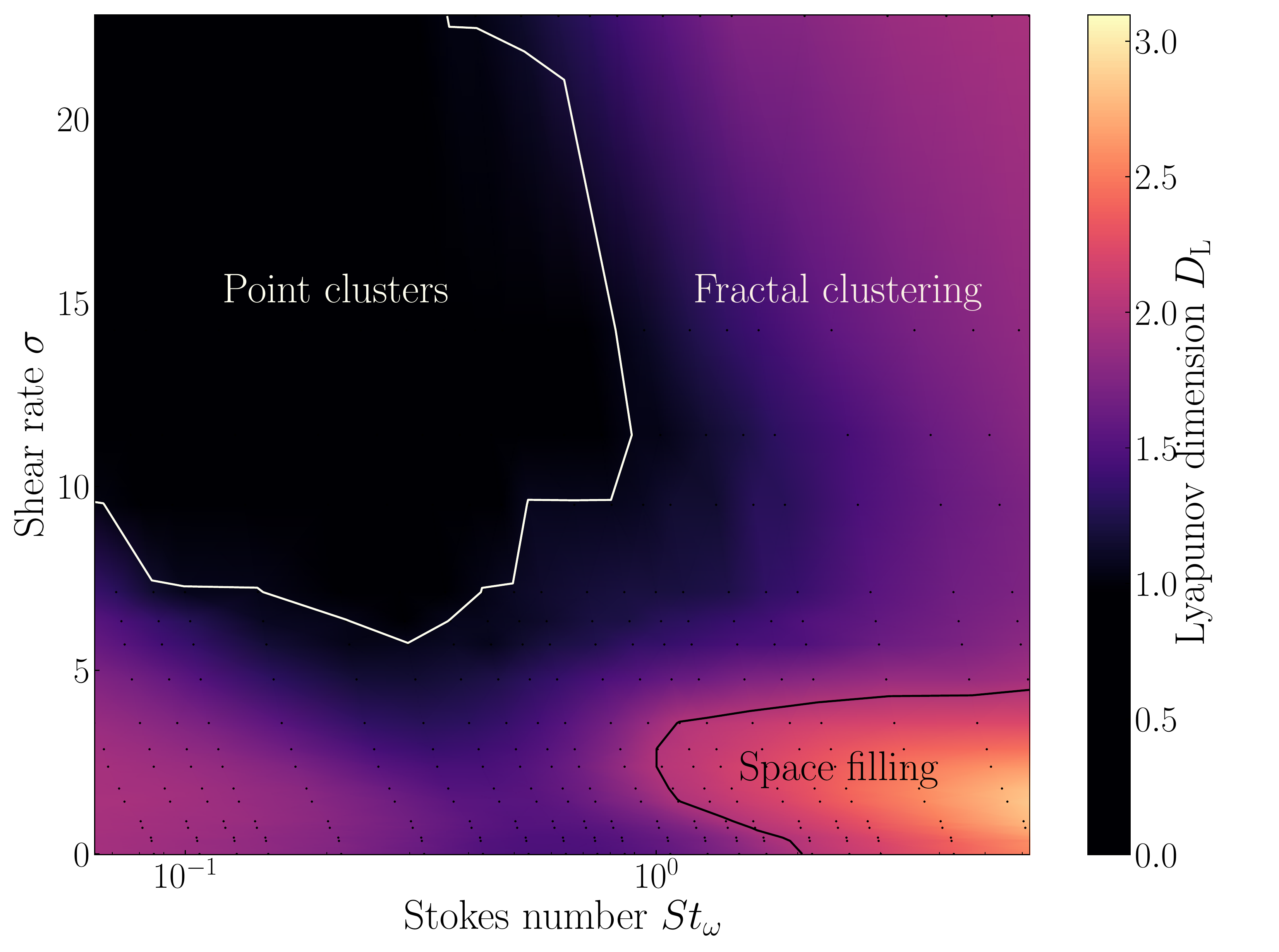}
    \caption{Phase diagrams of the shear rate parameter $\sigma$ vs the Stokes number $St_{\omega}$. The color code is for the Lyapunov dimension $D_{\rm L}$. Each black dot stands for the parameters that are actually simulated.}
    \label{fig:lyapunov_2D}
\end{figure}
Figure~\ref{fig:lyapunov_2D} shows the Lyapunov dimension $D_{\rm L}$ as a function of $\sigma$ and $St_{\omega}$. At low rotation rates and large Stokes numbers, the particles are space-filling and $D_{\rm L}>2$. They otherwise form fractal clusters where $0<D_{\rm L}<2$. At parameters values where the largest Lyapunov exponent is negative, we have $D_{\rm L}=0$ and particles concentrate in point clusters. It is clear from this phase diagram that rotation favours the clustering of heavy particles. When comparing it with the predictions from linear analysis of Fig.~\ref{fig:kida_analysis}, one recognizes similar behaviours, and in particular point clusters are found to occur at similar parameter values (fast-enough rotation and intermediate Stokes numbers).

\begin{figure}[htpb]
    \centering
    \includegraphics[width=\columnwidth]{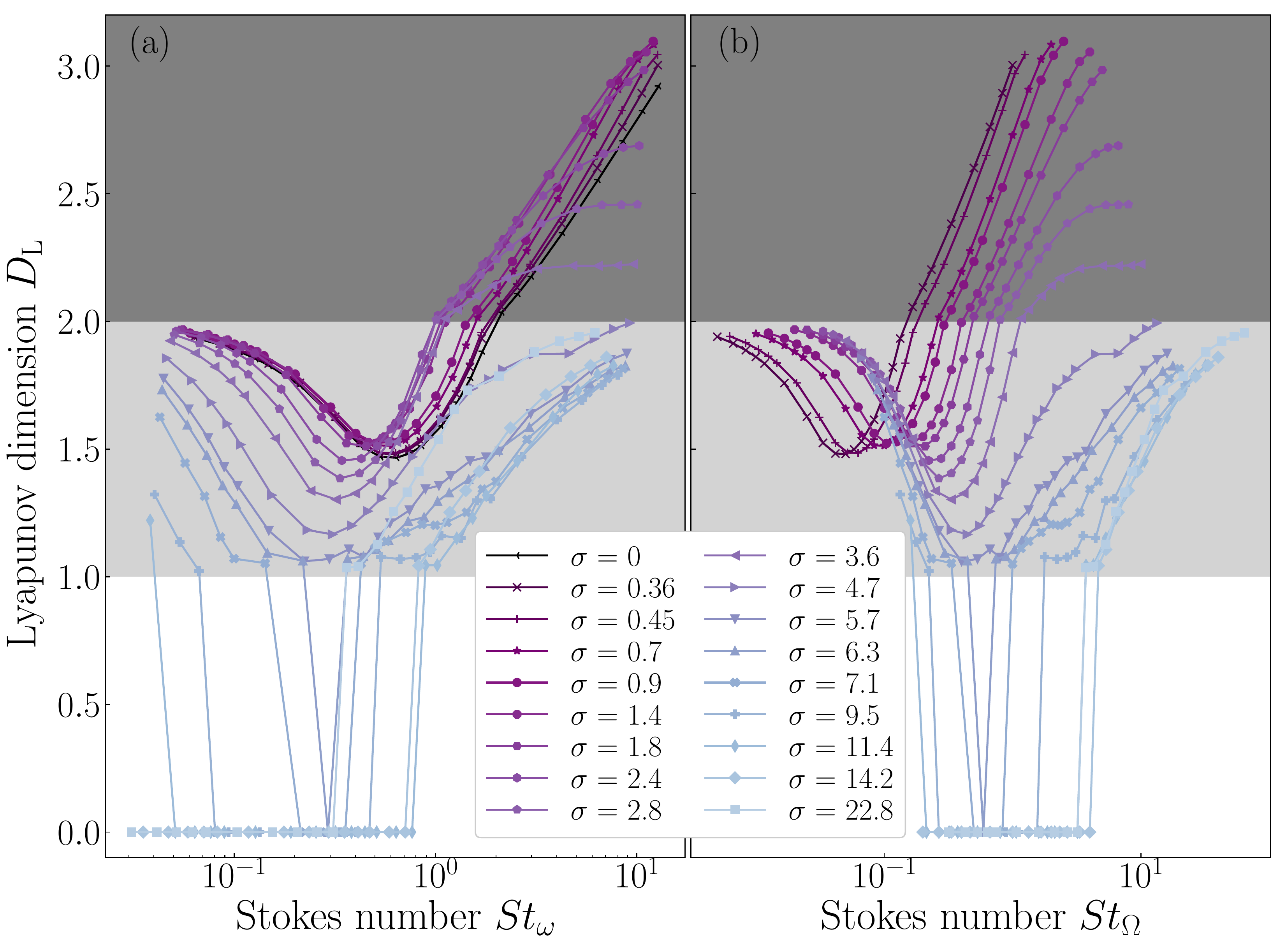}
    \caption{Lyapunov dimension $D_{\rm L}$ as a function of (a)~the ``turbulent'' Stokes number $St_{\omega}$  and (b)~the ``rotational'' Stokes number $St_{\Omega}$, for various values of the shear rate parameter $\sigma$ as labelled.}
    \label{fig:lyapunov}
\end{figure}
Figure~\ref{fig:lyapunov} shows the Lyapunov dimension, for various values of the shear rate parameter $\sigma$, this time as a function of the Stokes number. We compare the two different definitions introduced in Sec.~\ref{section:particles}, namely $St_\omega = \tau_{\rm p}/\tau_\omega$ (panel a) and $St_\Omega = \tau_{\rm p}\Omega$ (panel b). At slow rotations, small particles form fractal clusters, while larger ones fill space. At larger rotation rates, there is no space-filling regime and the Lyapunov dimension always remains below $d=2$. The ``turbulent'' Stokes number $St_{\omega}$ seems the most relevant to describe either small rotation rates or moderate particle response times.  For instance, at small to moderate values of $\sigma$, the strongest clustering (where $D_{\rm L}$ attains its minimum) is obtained for $St_{\omega}\approx 1$. Conversely, the asymptotics of large rotation rates and large particle response times seem to depend on $St_{\Omega}$ only. Indeed, the bottom-right curves shown in Fig.~\ref{fig:lyapunov}b almost collapse on the top of each other. In the limit $\sigma\to\infty$, turbulence becomes weaker and the turnover time $\tau_\omega$ increases. The relevant physics is then obtained by balancing particle inertia to the mean shear, whence a dependence upon $St_{\Omega}$ only. Such considerations could certainly be formalised in terms of fast and slow timescales allowing one to treat this asymptotics using averaging techniques (for example, through a WKB expansion). 

In order to further characterize the particle distribution, we have also studied higher-order statistics of their mass. Standard box-counting techniques rely on the mass (or number) of particles $m_\ell$ contained in a box of size $\ell$. The fractal nature of the particle distribution implies that the moments of $m_\ell$ behave as power laws: $\langle m_\ell^n \rangle \sim \ell^{\epsilon_n}$, where the exponents $\epsilon_n$ characterize the anomalous scaling typical of multifractal sets. More specifically, $\epsilon_n=(n-1)D_n+2$ where the $D_n$'s define the dimension spectrum of the attractor. They estimate how strongly a set of $n$ particles clusters.

\begin{figure}[htpb]
    \centering
    \includegraphics[width=0.9\columnwidth]{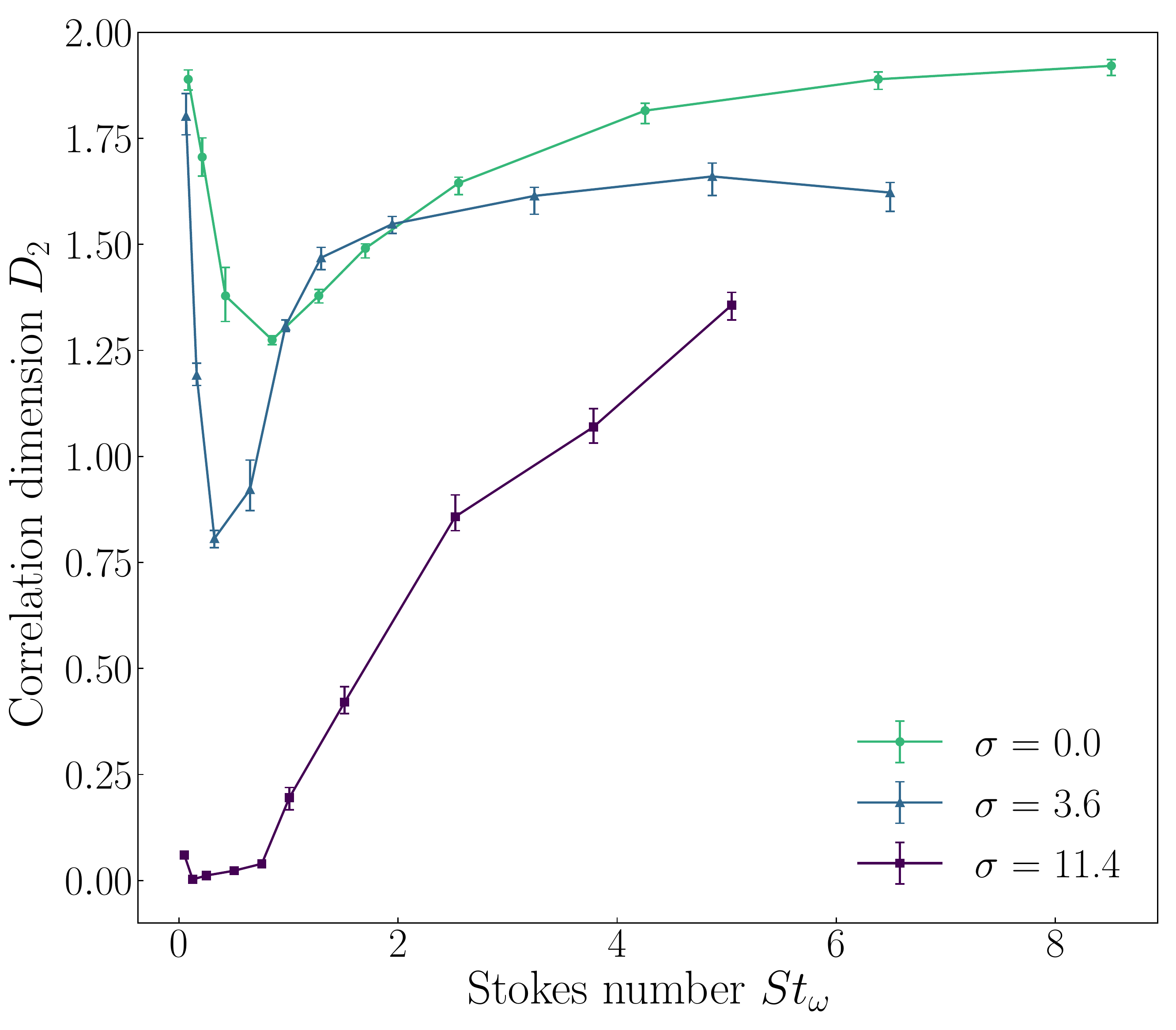}
    \caption{Correlation dimension $D_2$ as a function of the Stokes number $St_{\omega}$ for three values of the shear rate parameter $\sigma=\tfrac{3}{2}\Omega\tau_{\rm f}$.}
    \label{fig:correlation_dimension}
\end{figure}
We first focus on the correlation dimension $D_2$. It measures how to the probability $p_2(r)$ that two particles on the same attractor have a separation distance $<\ell$ depends on scale: Namely we have $p_2(r)\sim \ell^{D_2}$. When $D_2<2$, pairs of particles are more strongly concentrated than those uniformly distributed. We calculate $D_2$ using box-counting and measure the scaling exponent of $\langle m_\ell^2\rangle$ from the average local slope (logarithmic derivative) over a decade in length scale. Figure~\ref{fig:correlation_dimension} shows $D_2$ as a function of $St_\omega$ for three different values of the shear rate parameter $\sigma$. Error bars are here obtained from the maximum and minimum deviations of the local slope from its average. $D_2$ reaches a minimum at $St_{\omega}\approx 1$, independently of the rotation rate. In the absence of shear, the correlation dimension saturates to $D_2\approx2$ at large values of $St_{\omega}$, indicating that particles fill space. On the contrary, at larger values of $\sigma$, even the particles with large $St_{\omega}$ show fractal clustering. For $\sigma=3.6$, the dimension seems to actually saturate to $D_2\approx 1.6$. This result highlights once again the effect of rotation on the particle dynamics.

\begin{figure}[htpb]
    \centering
    \includegraphics[width=\columnwidth]{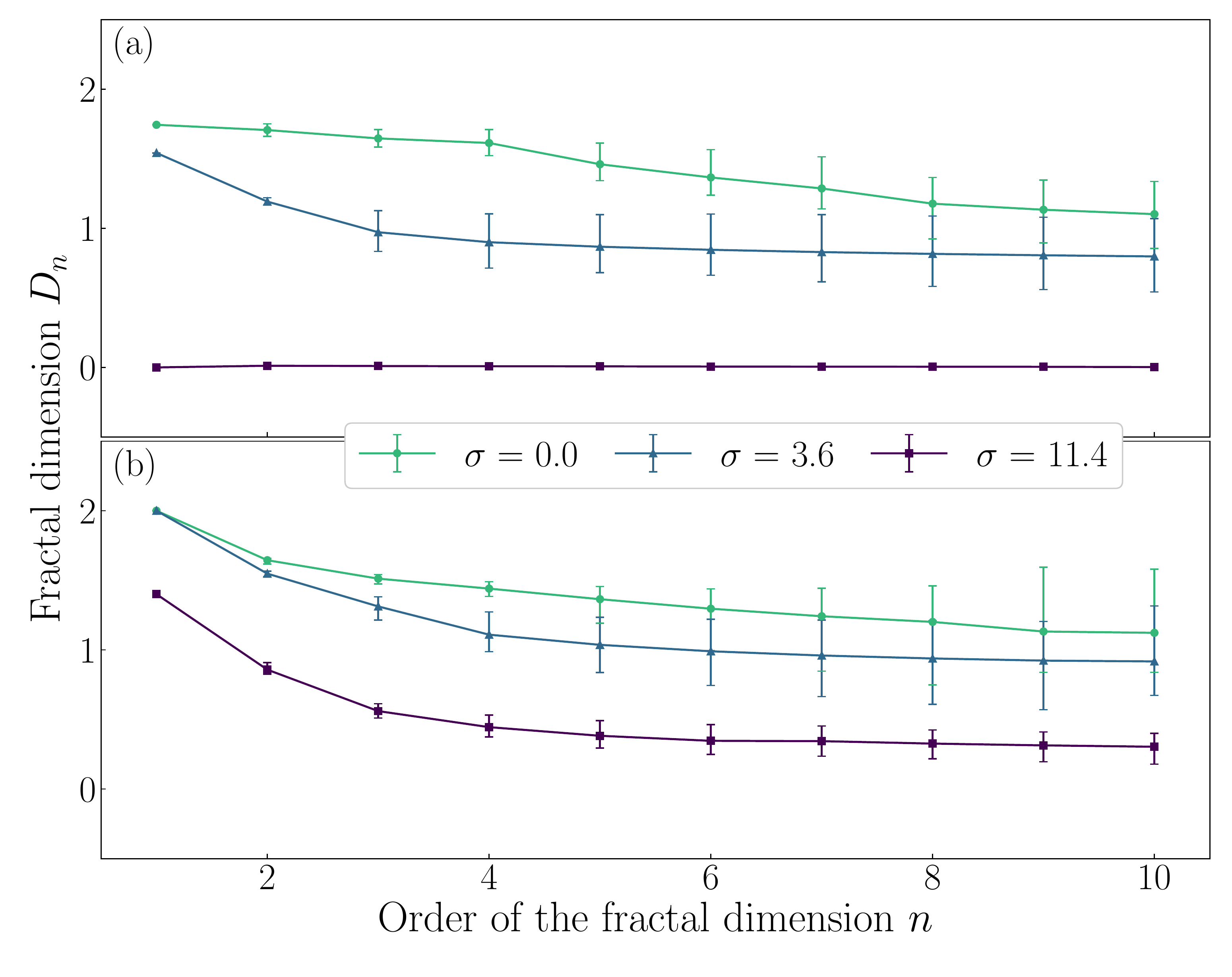}
    \caption{Fractal dimension $D_n$ as a function of the order of the fractal dimension $n$ for three values of the shear rate parameter $\sigma$ and Stokes number $St_{\omega}\approx0.2$ (a) and $St_{\omega}\approx2$ (b).}
    \label{fig:fractal}
\end{figure}

\begin{figure}[htpb]
    \centering
    \includegraphics[width=0.98\columnwidth]{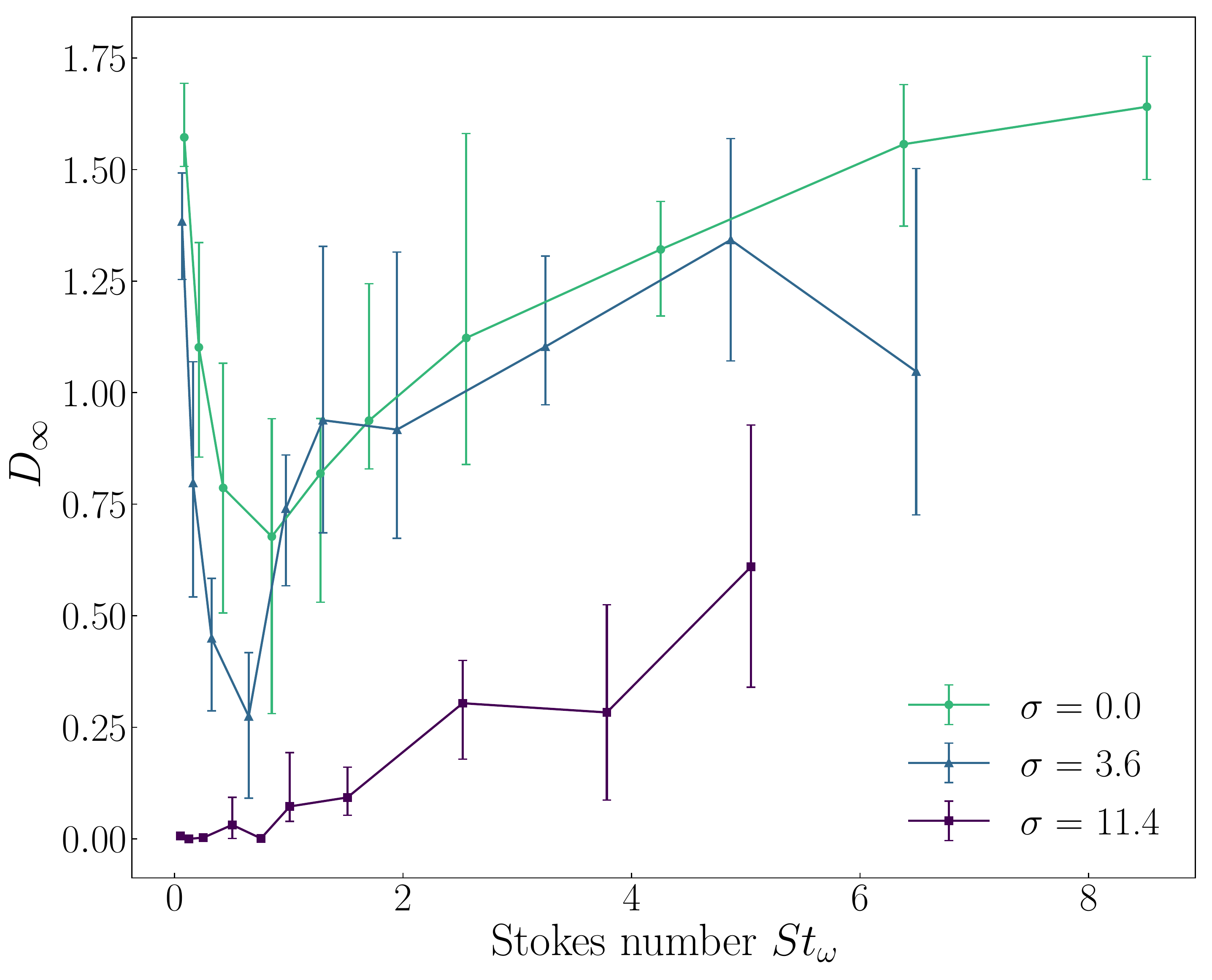}
    \caption{$D_{\infty}$ as a function of the Stokes number $St_{\omega}$ for different values of the shear rate parameter $\sigma$.}
    \label{fig:infinity}
\end{figure}

Figure~\ref{fig:fractal} shows the fractal dimensions $D_n$, as a function of the order $n$, for two values of the Stokes number $St_{\omega}\approx0.2$ (a) and $\approx2$ (b) and for the same three values of $\sigma$. In the case when particles form point clusters (Fig.~\ref{fig:fractal}a, $\sigma=11.4$), all fractal dimensions are equal to $D_n=0$. In the other cases, one clearly observes multifractal clustering with fractal dimensions $D_n$ that decrease with $n$, likely to tend to a limiting value $D_\infty$ at large orders. This behaviour is particularly interesting in the context of the gravitational collapse of dust. Large groups of particles would indeed interact as if distributed on a set of dimension $D_\infty$, significantly lower that the actual space dimension. Figure \ref{fig:infinity} shows $D_{10}$, which gives a fair approximation of $D_\infty$, for the various values of $\sigma$. The qualitative shape of variations as a function of $St_{\omega}$ is very similar to that of $D_2$ shown in Fig.~\ref{fig:correlation_dimension}. Still, despite large error bars, these measurements show that large-order dimensions can be very low over a rather wide range of Stokes numbers. In the absence of rotation ($\sigma=0$), one observes $D_\infty\approx 0.7$ at Stokes numbers for which clustering is maximal. Additionally, saturation to $D_\infty=2$ seems to occur at Stokes numbers significantly larger than those we considered here.
\begin{figure}[htpb]
    \centering
    \includegraphics[height=0.8\columnwidth]{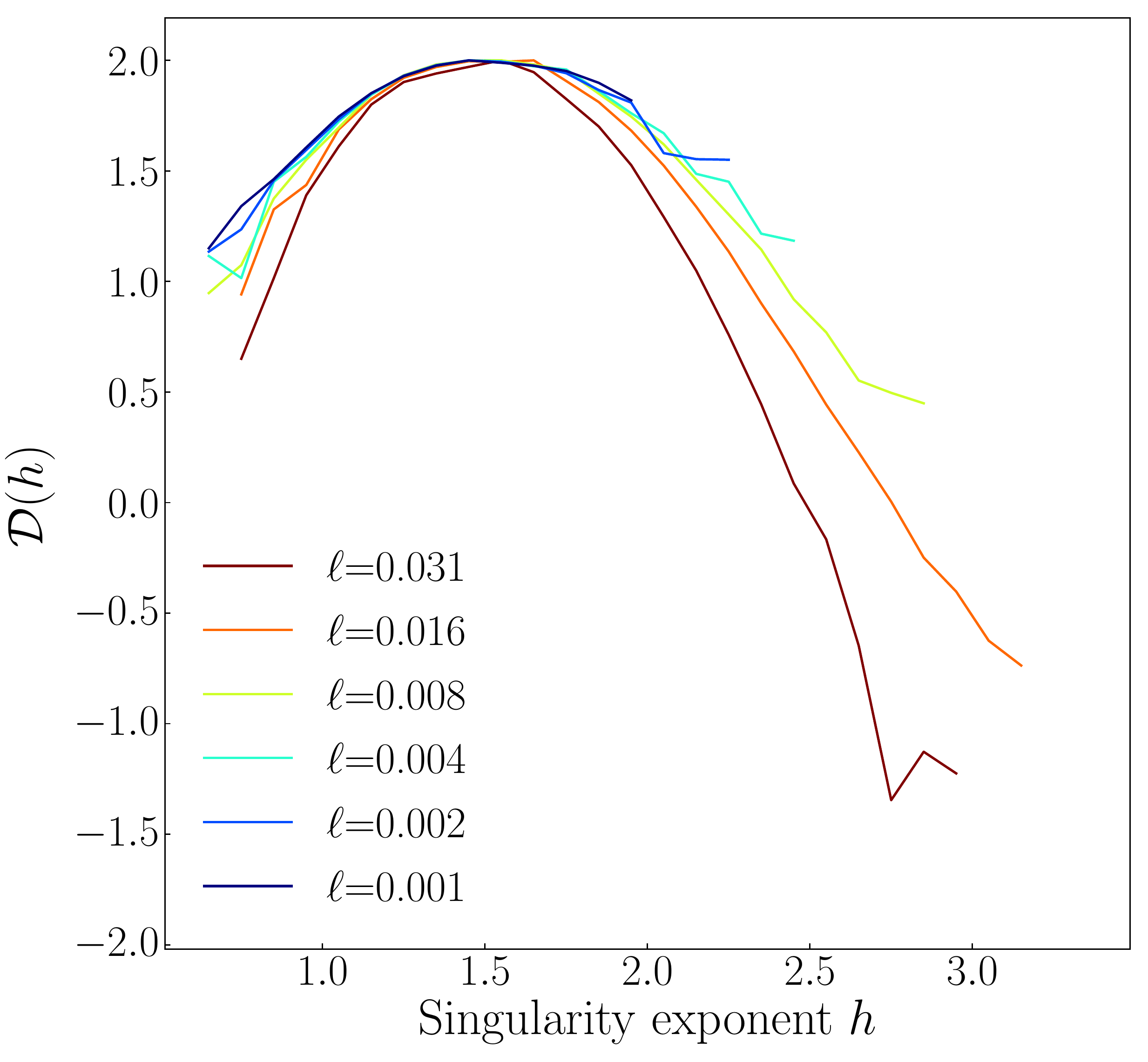}
    \caption{Dimension spectrum $\mathcal{D}(h)$ as a function of the singularity exponent $h$ for different values of the box size $\ell$ for $\sigma=3.6$ and $St_{\omega}=0.16$.}
    \label{fig:large_deviations_mass}
\end{figure}

Properties of the mass distribution can be equivalently described in terms of its large deviations. A multifractal distribution implies that the coarse-grained mass $m_\ell$ scales as $\ell^h$ on a fractal set of dimension $\mathcal{D}(h)$, that is with a probability $\propto \ell^{2-\mathcal{D}(h)}$. The scaling exponent $h$ is usually known as the singularity exponent, and $\mathcal{D}(h)$ is called the dimension spectrum. It is related to the fractal dimensions by a Legendre transform (see, \textit{e.g.}, \cite{PV87}): $D_{n+1}=\inf_h[hn+2-\mathcal{D}(h)]$. Figure~\ref{fig:large_deviations_mass} shows the dimension spectrum estimated from the probability distributions of $m_\ell$ at various scales $\ell$. All curves seem defined only for $h>h_{\rm min}\approx 0.7$ and $\mathcal{D}(h)$ has a finite limit when $h\to h_{\rm min}$. Using the Legendre transform that relates $\mathcal{D}(h)$ and $D_n$, one shows that this behavior is equivalent to having the fractal dimension saturating at large order to $D_\infty=h_{\rm min}$. The values we obtain from these two viewpoints are compatible with each other.

\section{Time evolution of point clusters}
\label{section:strong}
We have seen in the previous section that particle distributions develop a non-trivial, multifractal nature that can be used to quantify mass fluctuations. Clearly, this only applies to parameter values that fall in the ``fractal clustering'' phase of Fig.~\ref{fig:lyapunov_2D}. For values of the Stokes number and of the shear-rate parameter corresponding to the particles concentrating on point clusters, the mass distribution becomes trivial at long times: All fractal dimensions vanish and, equivalently, singularity exponents are restricted to $h=0$ with the associated dimension spectrum $\mathcal{D}(0)=0$. Nevertheless, in this point-clustering regime all the relevant physics occurs during transients.

\begin{figure}[htpb]
    \centering
    \includegraphics[width=\columnwidth]{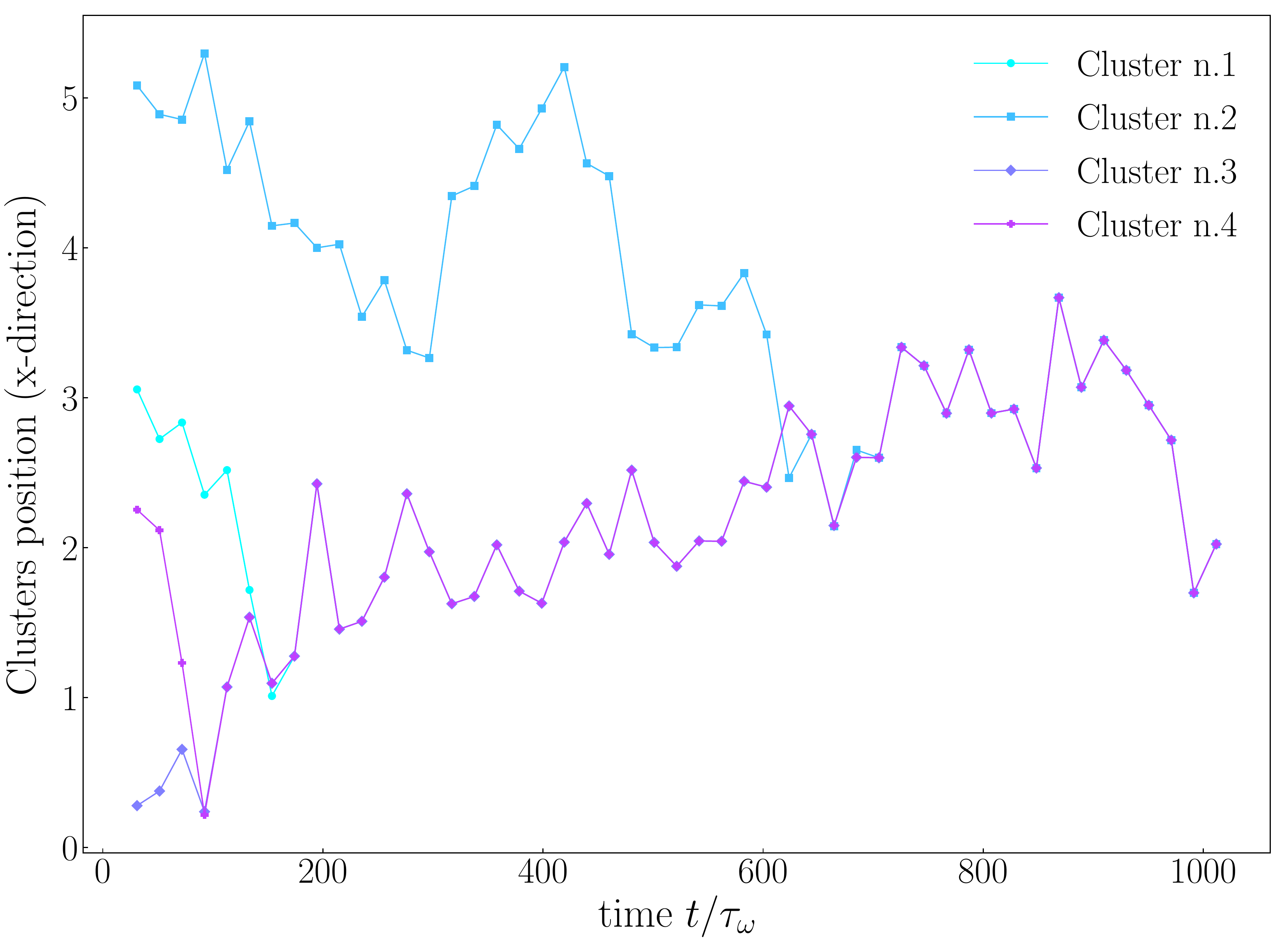}
    \caption{Position in the $x$-direction of particle clusters as a function of time for $\sigma=11.4$ and $St_{\omega}=0.19$.}
    \label{fig:Clusters_path}
\end{figure}

The Lyapunov exponent $\lambda_1$ that we measured in Fig.~\ref{fig:lyapunov_exp} gives an estimate of how fast a volume of particles converge to the center of a vortex. This tells us about the timescale of local concentration, but the convergence to a unique point cluster is actually ruled by other non-local mechanisms. Our simulations show that, before reaching a steady state (illustrated for instance~in Fig.~\ref{fig:movie}c), particles tend to concentrate in several anticyclones. These clusters then survive for long times (of the order of the vortex lifetime) before successively merging together and ending up in the same coherent structure. Such an evolution is illustrated in Fig.~\ref{fig:Clusters_path}, which shows how the  positions of the various particle clusters (here projected on the $x$ axis) evolve as a function of time. To define clusters, we have here divided the periodic domain $[0,2\pi[\times[0,8\pi[$ in $256\times1024$ square boxes. Those containing more than 10$\%$ of the total number of particles were then identified as clusters. In the specific case shown in Fig.~\ref{fig:Clusters_path}, four different clusters are found to be already formed after $t\approx40\tau_{\omega}$. We then track their time evolution and observe that they successively merge and finally form a unique point cluster at the end of the simulation.

\begin{figure}[htpb]
    \centering
    \includegraphics[width=\columnwidth]{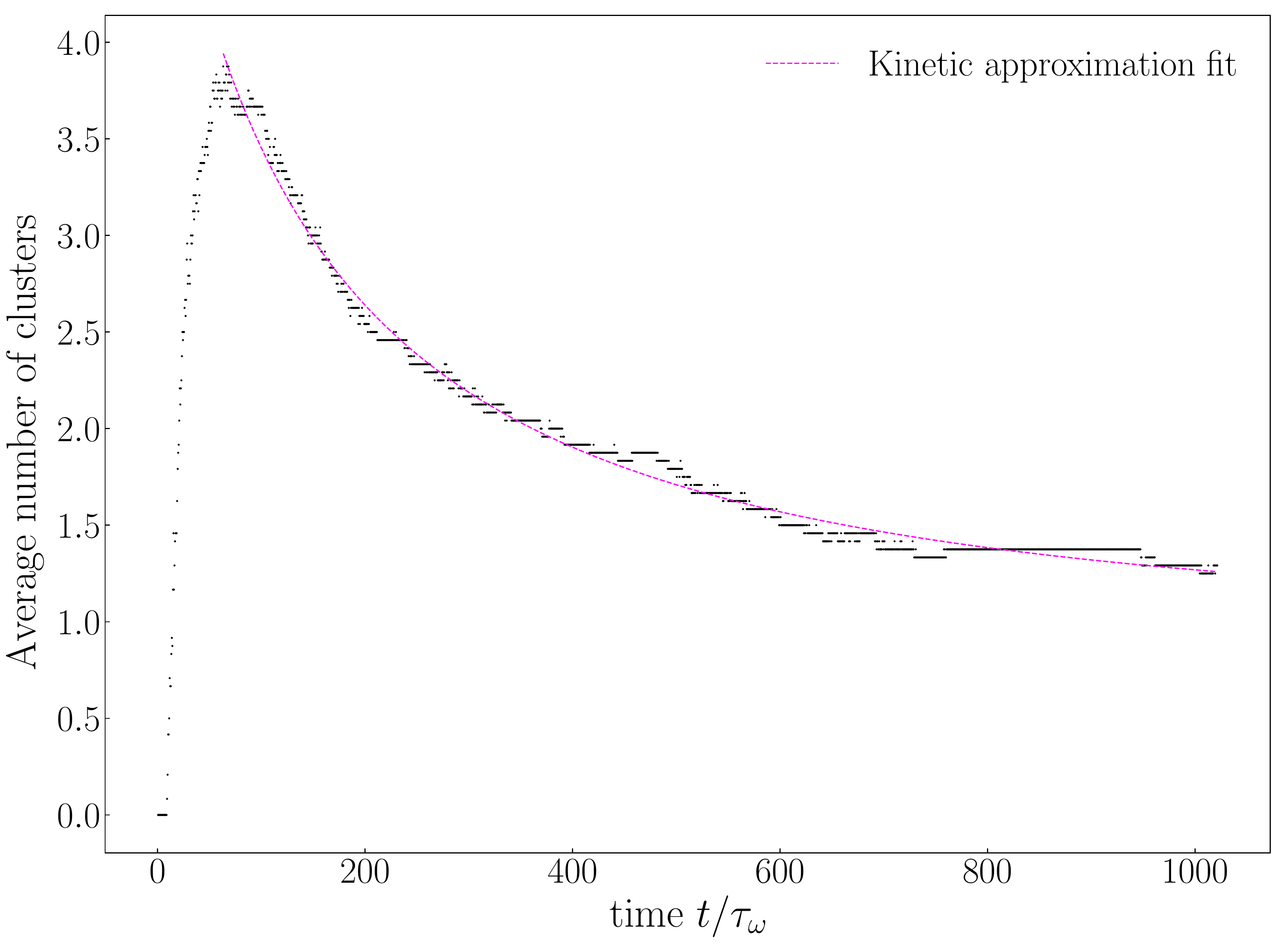}
    \caption{Average number of clusters as a function of time for $\sigma=11.4$ and $St_{\omega}=0.19$. The dashed line is a fit of the data obtained from the kinetic approximation~(\ref{eq:diff_clust}) with $\lambda\tau_\omega = 0.0014$ and $c = 0.81$.}
    \label{fig:Clusters_diff}
\end{figure}

The typical rate at which clusters merge together is given by the lifetime of vortices. This time is found to be $\tau_{\rm c}\approx225\,\tau_{\omega}$. It is much longer than the time $-1/\lambda_1\approx2.3\,\tau_\omega$ associated to particle concentration inside a given vortex core, but it is also much longer than all the physically relevant timescales of our problem. This separation of timescales suggests that cluster merger can actually be well approximated as a sequence of Poisson processes. Let us denote by $\bar{N}(t)$ the average number of clusters that are present in the domain at time $t$, where we assume that particles are uniformly seeded in the flow at $t=0$. Above considerations suggest that once clusters are formed, their average number approximately follows the kinetics
\begin{equation}
    \frac{{\rm d}\bar{N}}{{\rm d}t} \approx - \gamma \bar{N}(\bar{N}-1),
    \label{eq:kinetics}
\end{equation}
where the rate $\gamma$ is of the order of $\tau_{\rm c}^{-1}$. The above dynamics just assumes that between times $t$ and $t+{\rm d}t$, all pairs of clusters can independently merge with a probability $\gamma\,{\rm d}t$. Equation~(\ref{eq:kinetics}) integrates to
\be
\bar{N}\approx\frac{1}{1-c\,{\rm e}^{-\gamma\,t}}.
\label{eq:diff_clust}
\ee
We have repeated the cluster-tracking procedure that is explained above over an ensemble of 24 different realisations with the same physical parameters ($\sigma=11.4$ and $St_{\omega}=0.19$). Figure~\ref{fig:Clusters_diff} shows the time evolution of the average number of clusters then obtained. Data are well fitted by formula~(\ref{eq:diff_clust}) with $c = 0.81$ and $\lambda = 0.0014\,\tau_\omega^{-1} \approx 0.3\,\tau_{\rm c}^{-1}$ (dashed line). As anticipated, this rate seems ruled by the lifetime of anticyclones. However, because such measurements require a large number of realisations, we have not investigated how these numbers depend upon the shear parameter $\sigma$. Nevertheless, we expect stronger shears to deplete the survival of coherent structures, as for instance reported in~\cite{campana2022stochastic}. This implies that an increasing rotation rate could make the merger of point clusters a faster process. 

\section{Concluding remarks}
\label{section:disc}

In this paper we studied the clustering of heavy, dust particles in a turbulent flow with Keplerian rotation and shear. To do so, we performed 2D direct numerical simulations. The results have been analysed with tools borrowed from the study of nonlinear dynamical systems. The main findings are:
\begin{itemize}
    \item[--] The turbulent flow is substantially modified when increasing shear. Strong rotation favours anisotropies, with the preferential formation and survival of anticyclonic vortices.
    \item[--] Particles form point clusters in anticyclones or concentrate on fractal sets depending on both rotation and particle response time.
    \item[--] The mass distribution of dust particles is multifractal, with dimensions that saturate at large orders, and decrease with rotation.
    \item[--] Clusters form in a hierarchical manner and  merge one with the other in the course of time. This evolution occurs on timescales much longer than all other physical processes.
\end{itemize}
These results show that turbulence in protoplanetary disks is able to concentrate particles of various sizes. In the presence of fast rotation, these particles are rapidly and strongly clustered inside anticyclones. We expect these clusters to easily form planetesimals by gravitational collapse. We believe that planetesimals could form also from fractal sets, due to the important clustering strength between large groups of particles.  We give below more details on the importance and limits of our results on the astrophysical problem that was our primary motivation.

A first remark can be drawn from our extensive use of tools borrowed from nonlinear physics. This represents a rather powerful approach to tackle problems related to astrophysical disks dynamics. The new analysis tools that we could identify give original, innovative viewpoints on planetesimal formation. 

The effectiveness of such analysis tools does not depend on the details of the physical environment that we have considered. They remain of relevance in more complex situations. We indeed expect many aspects of dust clustering (point clustering in anticyclones, multifractal distributions, saturation of high-order dimensions, etc.) to be universal and independent of the turbulence-generating mechanisms (vertical shear instability, MRI, \ldots). In our model we considered a generic way to maintain a turbulent state, but different forcing terms could have been used to mimic the effects of specific instabilities. This would make our results more easily comparable with previous studies in astrophysics and using our tools to analyse such results could clearly provide new insights.

In our approach we used a fixed, white-noise forcing, thereby prescribing the inputs of energy and enstrophy to our turbulent flow. This allowed us to identify the key physical parameters for dust concentration, namely the rotation rate of the flow $\Omega$ and the particle response time to the flow dynamics $\tau_{\rm p}$. Most studies of planetesimal formation consider a fixed rotation rate.
At variance we have here changed it in a systematic manner in order to highlight how the combination of rotation and shear affect the properties of both the flow and the concentration of solid particles.

Performing such a systematic study of course required to simplify the physical setup and to neglect several mechanisms at play in protoplanetary disks.
First of all, we considered the gas to be in Keplerian rotation, as the dust particles are. With such a simplification, solids do not drift radially toward the star. Actually this average motion would have added a source of gas/dust velocity difference and thus modified the clustering of particles in anticyclones. Nevertheless, the drift timescale is known to be orders of magnitude larger than the rotation one, allowing to decouple the two processes. 
Another simplification was to neglect the back reaction of the dust onto the gas. Again, this is a second-order effect as the particles are strongly diluted. However, considering this coupling is needed to describe other planetesimal-forming processes, in particular the streaming instability which could be triggered locally in highly-concentrated regions. Finally, self-interactions between dust particles (\textit{e.g.}\/ collisions, gravity, etc.) have been disregarded. They should of course be included to follow dynamical evolution of particles after clustering. All these various aspects are kept for future work. 

As discussed in subsection \ref{subection:model}, we focused in this study on two-dimensional flows. Protoplanetary disks, despite being thin, have a vertical extension that is of great relevance for dust stratification. Solid particles indeed settle toward the mid-plane, on a timescale varying with the particle size. Numerous instabilities can also develop along the vertical direction, resulting in a turbulent state that can stir up again some solids to the disk surface. Full three-dimensional simulations should therefore be performed to address the question of the effect of vertical stratification on clustering. We nevertheless highlight again that at high rotation rate ($\Omega\gg1$) the flow is mainly two-dimensional, due to the formation of Taylor columns. We thus expect the formation of strong clusters of particles (lines in the core of columnar vortices)  to occur also in three dimensions, at least for large-enough rotation frequencies. Moreover, new disk observations~\citep{villenave2022} suggest that the vertical extent of dust is even smaller than previously thought. This requires solids to settle almost completely to the mid-plane, making their evolution nearly 2D from that moment on. 

Let us finally stress that protoplanetary disks are compressible, with the propagation of spiral density waves at large scales.  However, as already mentioned, we consider here sufficiently small scales to justify the incompressible approximation (see for instance~\citep{MFL15} for a comparison between compressible and incompressible turbulence in the shearing box). Addressing larger scales will of course require accounting for compressibility.

\section*{Acknowledgements}
This work received support from the UCA-JEDI Future Investments at the Universit\'e C\^ote d’Azur (funded by the French government, and managed by the National Research Agency,  ANR-15-IDEX-01).
This work was supported by the ``Programme National de Planétologie'' (PNP), ``Programme National de Physique Stellaire'' (PNPS) and ``Programme National de Physique et Chimie du Milieu Interstellaire'' (PCMI) of CNRS/INSU co-funded by CEA and CNES. HM acknowledges inspiring discussions that took place at the Aspen Center for Physics, which is supported by National Science Foundation grant PHY-1607611. FG and JB  would like to thank the Isaac Newton Institute for Mathematical Sciences, Cambridge, for support and hospitality (EPSRC grant No EP/R014604/1) during the programme ``Mathematical aspects of turbulence: where do we stand?'' where a part of this work was undertaken.

\bibliography{shearbox}

\end{document}